\documentclass[useAMS,usenatbib]{mnras}
\usepackage{newtxtext,newtxmath}
\usepackage[T1]{fontenc}

\DeclareRobustCommand{\VAN}[3]{#2}
\let\VANthebibliography\thebibliography
\def\thebibliography{\DeclareRobustCommand{\VAN}[3]{##3}\VANthebibliography}

\usepackage{graphicx}	% Including figure files
\usepackage{amsmath}	% Advanced maths commands
\usepackage{listings}
\usepackage{siunitx}
\usepackage{pifont}
\usepackage{subcaption}
\usepackage{hyperref}
\usepackage{lipsum}
\usepackage{acro}
\usepackage{physics}
\usepackage{booktabs}  
\usepackage{placeins}  
\usepackage{pdflscape}
\usepackage{float} 
\usepackage{multirow}
\usepackage{multicol}
\usepackage[capitalize]{cleveref}

\newcommand{\av}[1]{{\color{black}#1}}
\usepackage{tabularx}      % automatic column width (for Description)
\usepackage{xcolor}        % optional colour coding
\usepackage{threeparttable}% optional notes under table

\title[The radial acceleration relation from MIGHTEE and LADUMA]{MIGHTEE-HI / LADUMA: Investigating the link between baryons and dynamics with 130 resolved H{\sc i}-selected galaxies }

\author[V\u{a}r\u{a}\c{s}teanu et al.]{Andreea A. V\u{a}r\u{a}\c{s}teanu$^{1}$\thanks{email: \href{mailto:andreea.varasteanu@physics.ox.ac.uk} {andreea.varasteanu@physics.ox.ac.uk}}, Matt J.~Jarvis$^{1,2}$, Harry Desmond$^{4}$, Anastasia A.~Ponomareva$^{3,1}$, Tariq Yasin$^{1}$, 
\and
Michalina Maksymowicz-Maciata$^{5}$, Ian Heywood$^{1,6,7}$, Natasha Maddox$^{5}$, 
Andrew J. Baker$^{8}$, 
\and
Laurent Chemin$^{9}$, Martin Meyer$^{10}$, Danail Obreschkow$^{11}$, 
Kristine Spekkens$^{12}$, Natalia Stylianou$^{1}$, 
\and
Rohan G. Varadaraj$^{1}$, 
Marcin Glowacki$^{13, 14}$, Maarten Baes$^{16}$, Abhisek Mohapatra$^{15}$
\\
$^{1}$Oxford Astrophysics, Denys Wilkinson Building, University of Oxford, Keble Road, Oxford, OX1 3RH, UK \\
$^{2}$Department of Physics and Astronomy, University of the Western Cape, Robert Sobukwe Road, 7535 Bellville, Cape Town, South Africa \\
$^{3}$Centre for Astrophysics Research, School of Physics, Astronomy and Mathematics, University of Hertfordshire, College Lane, Hatfield, AL10 9AB, UK \\
$^{4}$Institute of Cosmology \& Gravitation, University of Portsmouth, Dennis Sciama Building, Portsmouth, PO1 3FX, UK\\
$^{5}$School of Physics, H.H. Wills Physics Laboratory, Tyndall Avenue, University of Bristol, Bristol, BS8 1TL, UK\\
$^{6}$Centre for Radio Astronomy Techniques and Technologies, Department of Physics and Electronics, Rhodes University, PO Box 94, Makhanda, 6140, South Africa\\
$^{7}$South African Radio Astronomy Observatory, 2 Fir Street, Black River Park, Observatory, Cape Town, 7925, South Africa\\
$^{8}$Department of Physics and Astronomy, Rutgers, the State University of New Jersey, 136 Frelinghuysen Road, Piscataway, NJ
08854-8019, USA\\
$^{9}$Université de Strasbourg, CNRS, Observatoire astronomique de Strasbourg, UMR 7550, 67000 Strasbourg, France\\
$^{10}$International Centre for Radio Astronomy Research (ICRAR), University of Western Australia, 35 Stirling Highway, Crawley, WA 6009, Australia\\
$^{11}$ICRAR, The University of Western Australia, 35 Stirling Highway, Perth, WA 6000, Australia\\
$^{12}$Department of Physics, Engineering Physics and Astronomy, Queen's University, Kingston, Ontario, K7L 3N6, Canada\\
$^{13}$Institute for Astronomy, University of Edinburgh, Royal Observatory, Edinburgh, EH9 3HJ, UK\\
$^{14}$Inter-University Institute for Data Intensive Astronomy, Department of Astronomy, University of Cape Town, Cape Town, South Africa\\
$^{15}$Department of Astronomy, University of Cape Town, Private Bag X3, Rondebosch 7700, Cape Town, South Africa\\
$^{16}$Department of Physics and Astronomy, Universiteit Gent, Proeftuinstraat 86 N3, B-9000 Ghent, Belgium}

\date{Accepted XXX. Received YYY; in original form ZZZ}

\pubyear{\the\year{}}

\begin{document}
\label{firstpage}
\pagerange{\pageref{firstpage}--\pageref{lastpage}}
\maketitle

% Abstract of the paper
\begin{abstract}
The baryonic Tully-Fisher relation (bTFR) and the radial acceleration relation (RAR) link the observed dynamics in galaxies to that expected from their baryonic mass distributions. The relations' small intrinsic scatters place strong constraints 
on galaxy formation models, dark matter properties and theories of modified dynamics, yet detailed measurements beyond the very local Universe 
remain limited. 
In this work, we use 130 purely H\,\textsc{i}-selected galaxies with both resolved H\,\textsc{i} kinematics and resolved baryonic mass profiles to measure the bTFR and RAR up to $z \approx 0.09$.
We measure a tight RAR with an acceleration scale $a_0 = (1.50 \pm 0.05) \times 10^{-10}~\mathrm{m\,s^{-2}}$ and an 
intrinsic scatter of $0.096 \pm 0.006$~dex, broadly consistent with previous results based on local samples. 

 We fit the bTFR in the `inverse' direction, conditioning on $M_{\rm bar}$ to mitigate H\,\textsc{i} flux-related selection effects,
 measuring a logarithmic slope of $0.27 \pm 0.01$ (corresponding to a forward slope of $3.72 \pm 0.16$), with vertical intrinsic scatter $\sigma_\perp \approx 0.05$~dex. 
 Fitting the more general $\delta$-family of MOND interpolating functions to the RAR, we infer a shape parameter $\delta = 4.10^{+1.4}_{-0.68}$, consistent with the value required by Solar System gravitational constraints and a null Wide Binary Test.  
 We find no significant redshift evolution in the RAR acceleration scale for our pure H{\sc i}-selected sample. On the other hand, the bTFR zero-point shows an apparent evolutionary trend that is strongly dependent on the fit direction: the traditional forward fit yields an $8.7\sigma$ preference for $z$ evolution, while for our fiducial inverse fit, this reduces to $3.4\sigma$, within $\approx 2\sigma$ of the RAR evolution constraint.
This suggests selection effects bias the forward fit; a careful consideration of such effects will be required in future endeavours to robustly measure the redshift evolution of dynamical scaling relations.
\end{abstract}


%%%%%%%%%%%%%%%%%%%%%%%%%%%%%%%%%%%%%%%%%%%%%%%%%%

%%%%%%%%%%%%%%%%% BODY OF PAPER %%%%%%%%%%%%%%%%%%

\section{Introduction} 
\label{intro}
How to interpret observational signatures of the missing mass or gravity is one of the central open questions in astrophysics. 
Such signatures include galaxy rotation curves: rather than declining at large radii as expected from Newtonian dynamics applied to the luminous matter alone, they remain approximately flat, implying the presence of an unseen mass component that dominates the outer regions of galaxies \citep{Rubin_1978, Bosma_1978, vandenBergh_2001, McGaugh_2004}.  
This observation is a key motivation for the $\Lambda$CDM cosmological model, in which galaxies form and evolve within massive dark matter haloes. Within this paradigm, the formation and evolution of galaxies are outcomes of a highly complex and stochastic interplay of processes: gas accretion, merger history, conversion of gas into stars and feedback from supernovae and active galactic nuclei. 
Despite this complexity, galaxies evince a high degree of regularity that manifests through tight dynamical scaling relations suggesting an intimate coupling between baryonic and dark matter components that is not trivially understood nor expected in the current $\Lambda$CDM framework. 
The baryonic Tully--Fisher relation (bTFR) is one such relation, connecting the total baryonic mass of a galaxy to its flat rotation velocity, $M_{\rm bar} \propto V_{\rm out}^s$, spanning over five magnitudes in mass with remarkably small scatter \citep{Tully_Fisher_1977, McGaugh_2000, McGaugh_2012, 
Ponomareva_2017, Harry_2017_bTFR, Lelli_2016b, Ponomareva_2021}. 
The radial acceleration relation (RAR) generalises the bTFR by linking the observed centripetal acceleration $g_{\rm obs}$, derived from rotation curves, 
to the acceleration predicted from the baryonic mass distribution 
alone, $g_{\rm bar}$, at every resolved radius within a galaxy. 
First measured by \citet{McGaugh_Lelli_2016} using the SPARC database of 175 
late-type galaxies with accurate H\,{\sc i} rotation curves and 
Spitzer 3.6\,$\mu$m photometry \citep{Lelli_2016}, the RAR 
reveals a tight relation between $g_{\rm obs}$ and $g_{\rm bar}$ across the full radial extent of galaxies.
At high accelerations, the RAR follows a one-to-one relation, 
implying that the observed baryons alone are sufficient to explain the observed dynamics. Below a characteristic acceleration scale of 
$a_{0} \approx 10^{-10}$\,m\,s$^{-2}$, the observed dynamics 
deviate significantly from the baryonic prediction, a trend typically attributed to the influence of dark matter. Remarkably, the intrinsic scatter of the RAR is extremely small---potentially consistent with zero \citep{McGaugh_Lelli_2016, Lelli_2017, rar_paper_2025}---making it the tightest known dynamical scaling 
relation governing the radial dynamics of late-type galaxies \citep{Desmond_2023, Stiskalek_2023}.

Efforts to reproduce the RAR within $\Lambda$CDM models have yielded mixed results. A variety of cosmological hydrodynamical simulations \citep{Wadsley_2017, Ludlow_2017, Tenneti_2017, Dutton_2019} and semi-empirical analytic models \citep{DiCintio_2016, Paranjape_2021, Li_2022, Desmond_MDAR} have 
been shown to produce a relation broadly resembling the RAR within the $\Lambda$CDM framework. However, their prescriptions differ from each other, and it remains unclear whether they can reproduce the full set of empirical properties of the RAR---its functional form, remarkably low intrinsic scatter, and the full diversity of rotation curves.
Its small intrinsic scatter and only marginal consistency with $\Lambda$CDM simulations make the RAR a powerful diagnostic for constraining galaxy formation models \citep{Desmond_MDAR} and may provide support for alternative theories of gravity, notably Modified 
Newtonian Dynamics (MOND; \citealt{Milgrom_1983, Sanders_1990}), which predicted this relation before it was empirically discovered \citep[e.g.][]{Milgrom_1983}.

Observationally, the RAR has been extended beyond rotationally supported spirals and irregulars to local dwarf spheroidals, pressure-supported early-type galaxies, ultra-diffuse galaxies in clusters \citep{Freundlich_2022}, galaxy 
groups and clusters \citep{Tian_2020, Chan_2020, Tian_2024, Bilek_2026}, weak lensing studies of isolated galaxies 
\citep{Brouwer_2021, Mistele_2024}, and even pulsars \citep{Yasin_Desmond_2026}. Recently, \citet{Julio_2025} extended the relation down to the very low mass regime ($10^{4} < M_{\rm bar}/M_{\odot} < 10^{7.5}$) using 12 nearby dwarf spheroidal galaxies. They find systematic deviations from the SPARC RAR at the 
lowest accelerations and disfavour a fully universal RAR in the smallest dwarf galaxy regime.

The evolution of the RAR with cosmic time has also begun to be explored: \citet{rar_paper_2025} used the MIGHTEE-HI galaxies from Data Release 1 \citep{Heywood_2024}, finding  tentative evidence for evolution up to $z\sim 0.08$. 
\cite{Ciocan_2026} report deviations from the local RAR with higher acceleration scale and intrinsic scatter, and significant evidence for evolution at the $30\sigma$level, using integral-field spectroscopy optical emission-lines observations of star-forming galaxies at $z \sim 0.3$--$1.5$.
Whether the intrinsic scatter of the RAR is consistent with zero, pointing to a fundamental relation, or not, as expected from galaxy formation in $\Lambda$CDM is also a crucial question.
Distinguishing between the two and quantifying the intrinsic scatter and acceleration scale $a_{0}$ hinges on the careful treatment of observational uncertainties and selection biases, and in particular the mass-to-light ratio \citep{rar_paper_2025}.

Many studies of the SPARC sample have either adopted fixed mass-to-light ratios \citep{Lelli_2017} or varied them as nuisance parameters when fitting the RAR galaxy by galaxy 
\citep{Li_2018, Chae_2020b, Chae_2021, Chae_2022}, finding intrinsic scatter in the RAR of $< 0.1$\,dex. \citet{Desmond_2023} performed a full joint Bayesian inference, mapping degeneracies among all parameters, and inferred an intrinsic scatter of $0.034 \pm 0.001$\,(stat) $\pm 0.001$\,(syst)\,dex.  The assumption of a spatially constant stellar mass-to-light ratio ($\Upsilon_{\star}$) is a persistent source of uncertainty in determining $g_{\rm bar}$. Near-infrared photometry (e.g.\ Spitzer 3.6\,$\mu$m) provides an effective dust-free tracer of the older stellar population and is less sensitive to variations in age and metallicity than optical bands \citep{Bell_2001, Meidt_2014, Norris_2014, Rock_2015}, but even in the near-infrared, $\Upsilon_{\star}$ varies radially within galaxies as stellar populations change from older, redder centres to 
younger, bluer outer discs \citep{Bell_2001, MacArthur_2004, Tortora_2010}. 
The RAR is also closely coupled to the baryonic Tully-Fisher relation, the tight relation between baryonic mass and rotation velocity \citep{McGaugh_2000, McGaugh_2012, Ponomareva_2017, Lelli_2016b, Ponomareva_2021}: in the deep MOND limit the bTFR follows directly from the RAR's asymptotic behaviour at low accelerations, with a slope of four. The two relations can thus be understood as manifestations of the same tight underlying coupling between baryonic and dark matter on different scales -- the RAR locally, at every galaxy radius, and the bTFR a single global point per galaxy.

In an earlier study with a sub-sample of 19 galaxies from MIGHTEE over the COSMOS field \citep{rar_paper_2025, Ponomareva_2026}, we demonstrated that incorporating these radial variations through resolved SED fitting yields a substantially tighter RAR than a constant mass-to-light ratio, showing that $\Upsilon_\star$ is not spatially constant even in the near-infrared and establishing radially resolved $\Upsilon_\star(r)$ as essential for reliable measurements of the relation. 
In this paper, we extend that analysis to both relations, using a larger sample of 130 H\,\textsc{i}-selected galaxies, drawn from the MIGHTEE 
\citep{Jarvis_2016} and LADUMA \citep{Blyth_Baker_2016} surveys. The deep multi-wavelength coverage over the MIGHTEE and LADUMA fields---spanning from the ultraviolet 
through the far-infrared---enables us to derive $\Upsilon_{\star}$ 
as a function of galactocentric radius, removing the need to treat it as a free or fixed parameter. Combined with H\,{\sc i} rotation curves from the MIGHTEE-H\,{\sc i} \citep{Heywood_2024} and LADUMA data, this allows us to measure both the 
baryonic and dynamical components of the RAR and the bTFR, for a homogeneous H{\sc i}-selected sample, to $z< 0.09$.
Importantly, our observations probe the low-acceleration regime where the gravitational effects of dark matter are most pronounced and where discriminating between $\Lambda$CDM predictions and alternative theories is most critical.

This paper is organised as follows. In Section~\ref{data} we 
describe the MIGHTEE-H\,{\sc i} and LADUMA data and our sample of 130 galaxies. 
In Section~\ref{methods} we detail the photometric measurements, 
resolved SED fitting, rotation curve extraction, and surface mass 
density modelling. In Section~\ref{results} we present the RAR and bTFR for our sample, and in 
Section~\ref{summary} we discuss our results and summarise our 
conclusions. Throughout the paper we assume a $\Lambda$CDM cosmology 
with $H_0 = 70$\,km\,s$^{-1}$\,Mpc$^{-1}$, $\Omega_m = 0.3$, and 
$\Omega_{\Lambda} = 0.7$. Unless otherwise stated, all logarithms 
are base 10.

\section{Data}
\label{data}
\subsection{MIGHTEE-HI}
\label{MIGHTEE-HI}
The MeerKAT International
GigaHertz Tiered Extragalactic Exploration survey \cite[MIGHTEE; ][]{Jarvis_2016}, is a medium-deep, medium-wide survey, carried out with the 64 dish MeerKAT radio telescope in South Africa. It provides simultaneous radio continuum \citep{Heywood_2021,Hale_2025}, spectral line \citep{Heywood_2024} and polarisation observations \citep{Taylor_2024} over $\sim 20$\,deg$^2$ area covering the best studied extragalactic deep fields: COSMOS, XMM-LSS, Extended Chandra Deep Field South (ECDFS) and ELAIS-S1. 

The MIGHTEE-H{\sc i} emission project within MIGHTEE represents one of the first deep, 
medium-wide interferometric surveys for neutral hydrogen. The Early Science spectral line products are described in detail in \cite{Maddox_2021}.
Our previous science results \citep{rar_paper_2025} leveraged the H{\sc i} DR1 products over the COSMOS field \citep{Heywood_2024}; in this work we use these MIGHTEE-HI deep spectral line observations in COSMOS and new data covering 10.8\,deg$^2$ of the XMM-LSS field (Heywood et al. in prep).
Both these fields contain spectral-line coverage comprising two interference-free regions of MeerKAT L-band: L2 ($0 < z_{\rm HI} < 0.10$; 1290–1520 MHz) with 26 kHz channels (5.5 km s$^{-1}$ at $z=0$) used here, and L1 ($0.23 < z_{\rm HI} < 0.48$; 960–1150 MHz) with 104.5 kHz channels. Both sub-bands were also imaged at three different spatial resolution settings (for a detailed description see \citealt{Heywood_2024}). For this work we use the high angular resolution data with FWHM$\sim 12$ arcseconds.

\subsection{LADUMA}
\label{laduma}
We supplement our MIGHTEE sample with additional galaxies from the LADUMA field (Looking at the Distant Universe with the MeerKAT Array survey; \citealt{Blyth_Baker_2016}). 
LADUMA is an ultra-deep, untargeted 21\,cm HI survey conducted with the MeerKAT array, with a single pointing centred at $\alpha \approx 03^{\rm h}32^{\rm m}30^{\rm s}$, $\delta \approx -28\degr07\arcmin57\arcsec$ (J2000) in the Chandra Deep Field South (CDFS). 
The MeerKAT data in this case correspond to LADUMA's first 130\,hr of on-source L-band observations, which have been processed as described by \citet{kazemi_2025} - see also Kazemi-Moridani et al., in prep. The resulting data cubes have a finer angular resolution (the synthesized beam is nearly circular and in the range $7-8\,{\rm arcsec}$) and a coarser spectral resolution ($104.52\,{\rm kHz} \leftrightarrow 22\,{\rm km\,s^{-1}}$ after rebinning) relative to the MIGHTEE data products. All analysis (as for the MIGHTEE sources) has been done after correction for primary beam attenuatation.
Rotation curves are extracted using the same procedure as MIGHTEE galaxies (Section~\ref{HI-kinematics}). 

\subsection{Multi-wavelength data}
\label{multiwavelength_data}
\subsubsection{Optical and near-infrared}

For this work, we leverage the wealth of deep multi-wavelength data over the COSMOS, XMM-LSS and CDFS fields, which are crucial for stellar mass measurements.

For COSMOS and XMM-LSS fields, we use optical $u$-band photometry sourced from the Canada-France-Hawaii Telescope CFHT \citep{Cuillandre_2012}, while the \textit{$griz$} photometry is provided by the HyperSuprimeCam Subaru Strategic Program (HSC; \citealt{Aihara_2018}). 
For the near-infrared \textit{$YJHK_{s}$} photometry, we rely on the UltraVISTA Data Release 6 imaging \citep{McCracken_2012} and VIDEO (VISTA Deep Extragalactic Observations; \citealt{Jarvis_2013}).
The 5$\sigma$ depths of the optical data are in the range 25-27\,mag (AB) for a 2~arcsec aperture \citep[see Table~1 in][for more information]{Adams_2023}; thus they are significantly deeper than the data usually available for photometric measurements of relatively low-redshift galaxies.
In addition, the imaging has broadly comparable seeing ($\sim$ 0.8$\arcsec$), reducing the need for significant band-to-band aperture corrections.

For CDFS, we combine optical imaging from the VST Optical Imaging of the CDFS and ELAIS-S1 fields survey (VOICE; \citealt{Vaccari_2016}) with HSC where available. We take VOICE $u$, $g$, and $r$ \citep{Varadaraj_2023}, and complement with HSC imaging in $g$, $i$, and $z$, which is typically deeper in these bands. Near-infrared $YJHK_{\rm s}$ photometry is obtained from VIDEO imaging over CDFS.
The benefits of using these data is that all three fields lie at high Galactic latitude, minimising Milky Way foreground extinction.

\subsubsection{Mid-infrared}
We include mid-infrared photometry from the \textit{Spitzer} Space Telescope. Specifically, we adopt IRAC (Infrared Array Camera) 3.6 and 4.5~$\mu$m photometry \citep{Mauduit_2012} with an angular resolution of $\approx$ 2 arcseconds, and
24 $\mu$m data from the Spitzer Spitzer Wide-area InfraRed Extragalactic (SWIRE; \citealt{Lonsdale_2003}) survey with 6 arcsec angular resolution.

\subsubsection{Far-infrared}
Far-infrared photometry is taken from the Herschel Extragalactic Legacy Project (HELP; \citealt{Shirley_2021}\footnote{\url{https://herschel.sussex.ac.uk}}), which homogenises \textit{Herschel Space Observatory} data across our survey fields. We use imaging from the
Photodetector Array Camera and Spectrometer (PACS; \citealt{Poglitsch_2010}) at 100 and 160\,$\mu$m, with beam sizes of 6.7 and 11\,arcsec (FWHM) respectively, and from the Spectral and Photometric Imaging Receiver (SPIRE; \citealt{Griffin_2010}) at 250, 350, and 500\,$\mu$m, with beam sizes of 18, 25, and 36\,arcsec (FWHM), respectively. 
Together with the MIPS 24\,$\mu$m imaging described above, these bands sample the peak of the thermal dust emission and extend the SED coverage from the near-infrared to $\sim$500\,$\mu$m. The primary motivation for including far-infrared data is to constrain the total star formation rate (SFR) of each galaxy \citep{Tudorache_2026, Ramaiya_2026};  we use it to estimate total molecular gas masses via the \citet{Tacconi_2018} depletion-time scaling relation which enter the baryonic mass budget of the RAR and bTFR. 

\subsection{Sample selection} 
We construct an H{\sc i}-selected sample suitable for resolved kinematic analysis from the MIGHTEE H{\sc i} imaging in the COSMOS and XMM-LSS fields, supplemented by additional galaxies drawn from the LADUMA survey.
For the MIGHTEE component, in addition to the original RAR sample of 19 galaxies described in \cite{rar_paper_2025}, we include an additional 9 galaxies from the new MIGHTEE H{\sc i} source catalogue over the COSMOS field (see \citealt{Maksymowicz-Maciata_2026} for details).
We also include the galaxies catalogued in the 10.8\,deg$^2$ of the  XMM-LSS field, which comprise MIGHTEE-HI DR2 (Heywood et al. in prep.). These were identified using the same source finding method detailed in \cite{Maksymowicz-Maciata_2026}, which results in 96 H{\sc i} resolved galaxies out to $z\sim 0.09$. 
We exclude systems with bright foreground stellar contaminants in the 
optical bands that prevent reliable photometric and resolved baryonic mass 
modelling from both subsamples, leaving a total of 124 MIGHTEE and 6 LADUMA galaxies. We note that this does not bias our H{\sc i}-selected sample in any way.
The systemic velocities of our sample range from $V_{\rm sys}\approx 1720\,\mathrm{km\,s}^{-1}$ to $V_{\rm sys}\approx 25556\,\mathrm{km\,s}^{-1}$, with a mean value of around $14625\,\mathrm{km\,s}^{-1}$, with fewer than $6 \%$ of our galaxies below $z \approx 0.02$ ($cz \approx 6000\,\mathrm{km\,s}^{-1}$). 
Peculiar velocities are typically of the order of $300\,\mathrm{km\,s}^{-1}$; therefore, they represent only very small fractions of the recession velocities, which makes their effect a sub-dominant source of uncertainty in the stellar and baryonic masses for the majority of our sample \citep{Tully_2016, Richard_2026}.
We require that all galaxies have at least three resolution elements in the H{\sc i} data along their major axes. Full details of the selection and automated kinematic modeling can be found in  \cite{Ponomareva_2021, Ponomareva_2026} and \cite{rar_paper_2025}. 
Out of this initial selection, we further require those galaxies to have inclinations between $30^\circ < i < 80^\circ$; the lower limit ensures a meaningful $\sin(i)$ deprojection of velocities, and the upper limit excludes near edge-on systems where dust extinction and line-of-sight blending is prominent, which results in flattening of the inner slope of the rotation curves, as shown in various works \citep{Giovanelli_2004, Rhee_2004}.

\begin{figure}
    \centering
    \includegraphics[width=0.5\textwidth]{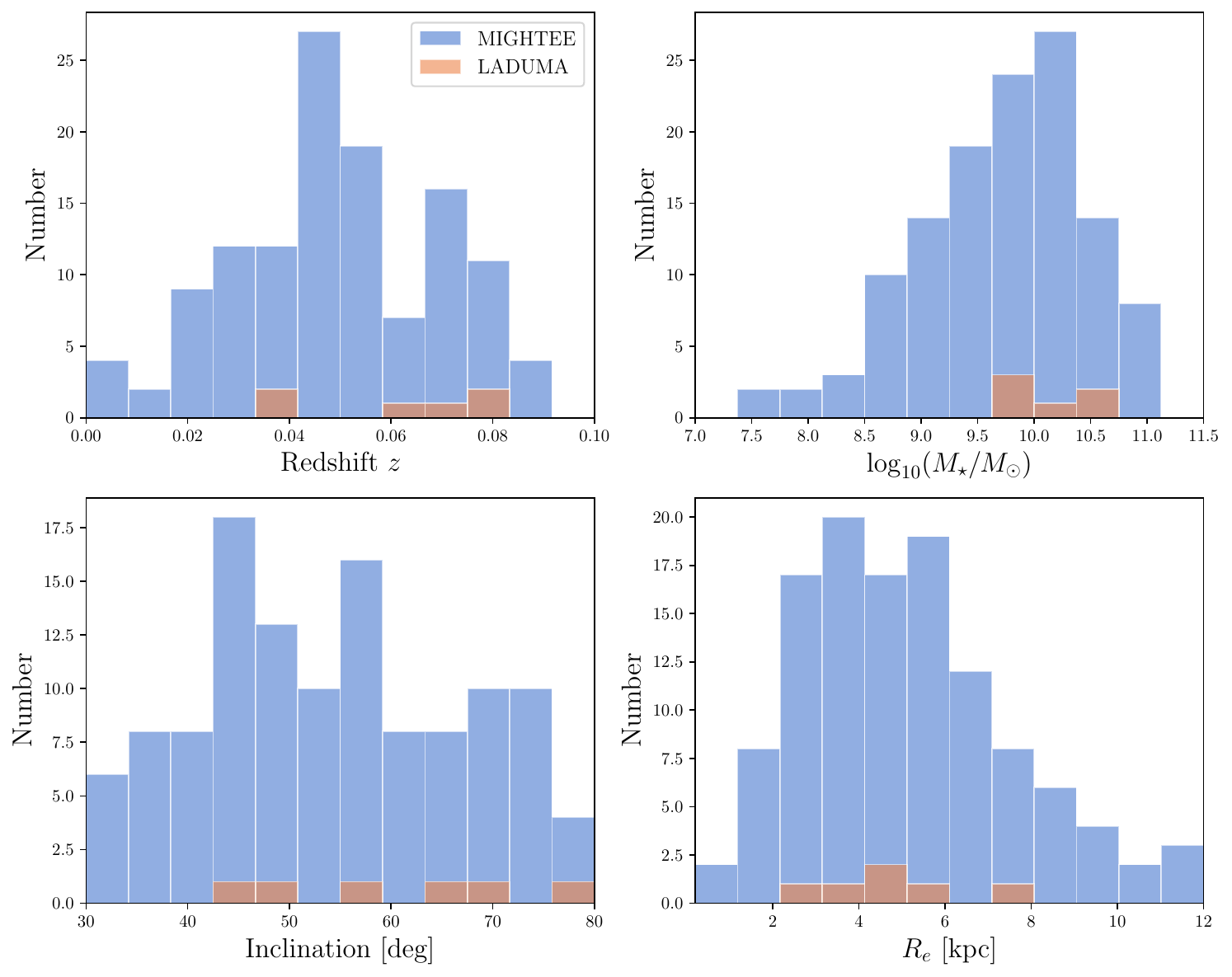}
    \caption{Distribution of physical properties for the galaxies in our sample. From top-left to bottom-right: redshifts, stellar masses, inclinations and effective radii in the near-infrared H band ($1.65 \mu m$).}
    \label{fig:sample_properties}
\end{figure}

The inclination angle is derived from the axis ratio measured in the optical $G-$ band, which traces the extended stellar disc and star forming regions, thus providing a good proxy for the geometry of the H\,\textsc{i} disc \citep{Ponomareva_2026}. 
We use the standard relation:
\begin{equation} 
    \cos^2(i) = \frac{(b/a)^2 - q_0^2}{1 - q_0^2},
    \label{eq:inclination}
\end{equation}
where $b/a$ is the observed minor-to-major axis ratio and $q_0$ is the 
intrinsic axis ratio of an edge-on disc (see \citealt{Tully_Fisher_1977}). 
Values of $q_0$ for late-type discs typically lie between 0 (thin disc) and 0.4 \citep{Fouque_1990, Weijmans_2014}. 
Given our use of a kinematic tracer (H{\sc{i}}) that is expected to be tightly confined to a galaxy's plane, we adopt the thin-disc assumption $q_0 = 0$, and verify that alternative choices change the derived inclinations by $\lesssim 5^\circ$, well within our quoted uncertainties.
The resulting sample comprises of 130 galaxies with high-quality, resolved rotation curves, up to $z = 0.09$. Combined with the deep multi-wavelength photometry across 18 bands from the optical to the 
far-infrared described in Section~\ref{multiwavelength_data}, this 
H\,\textsc{i}-selected dataset is uniquely suited to investigate the 
RAR and bTFR with a homogeneous, H\,\textsc{i}-mass-selected, sample. 
The distributions of redshift, 
stellar mass, inclination, and effective radius across the sample
is shown in Figure~\ref{fig:sample_properties}. \av{Owing to its depth, LADUMA adds galaxies at the high-redshift end of our sample ($z > 0.06$), where MIGHTEE contains relatively fewer objects.}

\section{Methods}
\label{methods}
\subsection{Photometry}
\label{photometry}
To extract photometry across all 18 wavelengths from optical to far-infrared, contaminant sources must first be masked in each image.  For this purpose, we developed a custom photometry workflow that makes use of {\sc sep} \citep{Barbary2016}, a Python implementation of Source Extractor (SExtractor \citealt{Bertin_1996}) with source extraction and photometry, to detect and mask contaminant sources in each photometric band prior to surface brightness fitting.

Source detection proceeds in two stages. An initial extraction without deblending defines the central galaxy's footprint as a single connected region. A second extraction with deblending then produces individual ellipse measurements for all detected sources. Sources outside the galaxy footprint are masked using a two-tier approach adapted to their morphology: extended sources are masked via their segmentation footprints directly, while compact and point-like sources are masked with brightness-scaled ellipses, where brighter sources receive the full aperture and fainter sources receive progressively smaller masks. Foreground stars projected onto the galaxy are identified with DAOStarFinder \citep{Stetson_1987} and masked only if they satisfy both a flux concentration criterion (to distinguish genuine stars from \ion{H}{II} regions or spiral-arm knots) and a brightness threshold. The central bulge region is protected from star masking to preserve the galaxy's core light profile. The result is a clean, background-subtracted image suitable for model fitting.
All images are then inspected by eye to identify galaxies that are contaminated by stray light from nearby bright objects, which are subsequently removed from the sample, and to check that the procedure does not over-subtract or misidentify regions within the H{\sc i} galaxy and foreground objects.

\subsubsection{Non-parametric photometry}
\label{direct_photometry}
We measure photometry directly in elliptical annuli, without imposing any functional form for the surface brightness profile. This non-parametric method aims to capture real deviations from smooth radial profiles such as spiral arms, bar structure etc., making it more suitable for galaxies with irregular and complicated morphologies that a smooth S\'ersic model cannot capture.

For each galaxy, we define concentric elliptical rings using the axis ratio and position angle measured from the $G$-band source detection and/or S\'ersic fit, choosing an annulus width of $1.5 \times \mathrm{FWHM}$ to ensure that adjacent annuli sample independent resolution elements. Within each annulus, we sum the background-subtracted pixel values from the masked image to obtain the total flux. The computation of photometric uncertainties on the total fluxes is described in Section~\ref{parametric_photometry}.

We apply Galactic extinction corrections to all photometric bands using Schlegel dust extinction maps \citep{Schlegel} assuming an $R_V$ = 3.1 extinction curve. 
 For the optical bands, we compute extinction coefficients using the \citet{Fitzpatrick_99} parameterisation via the Python package \textsc{extinction}. In the near-infrared, we adopt the extinction coefficients from the VISTA survey technical documentation.

\subsubsection{Parametric light profiles}
\label{parametric_photometry}
As a complementary approach to the non-parametric photometry described above, we also fit S\'ersic surface brightness profiles with {\sc AstroPhot} \citep{Stone_2023}. 
\textsc{AstroPhot} constructs a 2D model of the galaxy's surface brightness distribution, convolves it with the appropriate PSF, and fits to the observed image via chi-squared minimisation via a Levenberg–Marquardt optimiser, with per-pixel uncertainties incorporated through variance maps. We note that its likelihood treats pixels as independent, so the formal parameter uncertainties do not capture correlated noise and may lead to underestimates of parameter uncertainties. To account for this effect, we estimate flux uncertainties by measuring the scatter in aperture fluxes placed in blank regions of each image, which naturally captures the effect of correlated noise within that specific aperture. 
We fit both single and double S\'ersic models to each band simultaneously  with model selection guided by the Bayesian Information Criterion (BIC).
This approach allows us to recover intrinsic structural parameters---effective radius $R_e$, S\'ersic index $n$, axis ratio $q=b/a$, and position angle $PA$---as well as total fluxes. 
The fitting is performed on the tangent-plane projection in arcsecond coordinates, with masked pixels excluded from the likelihood evaluation. Following our previous work, we fix the position angle and ellipticity to the values measured in the $G$-band for all the other optical and near-infrared bands. 
For lower-resolution mid- and far-infrared bands, we likewise adopt the optical reference geometry by fixing the position angle and ellipticity. For galaxies resolved in Herschel/PACS we additionally fix all geometric parameters and fit only the intensity $I_e$. At the longest SPIRE wavelength  ($500 \mu$m), where all our galaxies are unresolved, we perform point-source photometry.

\subsubsection{SED fitting}
\label{sed_fitting}

We measure total stellar masses through spectral energy distribution (SED) fitting using \textsc{Bagpipes} (Bayesian Analysis of Galaxies for Physical Inference and Parameter Estimation; \citealt{Carnall_2018}), which compares observed photometry to synthetic SEDs generated from the \citet{Bruzual_Charlot_2003} stellar population synthesis models assuming a \citet{Chabrier_2003} initial mass function (IMF). \textsc{Bagpipes} operates under the assumption of energy balance, whereby the emitted dust luminosity equals the energy attenuated from the UV–optical spectrum.
We tested several parametric star formation history (SFH) models available in \textsc{Bagpipes}, including exponential, delayed exponential, log-normal, and double power-law forms. The choice of SFH does not significantly affect the recovered stellar masses; indeed, for most MIGHTEE-H{\sc i} galaxies no single SFH model is statistically preferred over the others \citep{Tudorache_2026}. We therefore adopt the log-normal SFH for all fits. Dust attenuation is modelled using the \citet{Calzetti_2000} law with A$_V$
 allowed to vary between 0 and 4\,mag; we verified that the choice of extinction law does not affect our results. Redshifts are fixed to the spectroscopic values derived from the H\,\textsc{i} emission line. The total stellar masses and their uncertainties are derived from the posterior distributions, taking the median of each marginalised posterior as the best-fit value; we note that these do not account for systematic uncertainties arising from the choice of IMF or SFH, and we adopt a minimum flux uncertainty of 5\% in the optical and near-infrared bands, and 20\% in the far-infrared. These uncertainties accommodate slight variations in the photometric zero points and model-data mismatch between the theoretical synthetic stellar population template set and the observations.
 The inclusion of near-infrared photometry is particularly valuable for robust stellar mass estimation, as these bands are minimally affected by dust attenuation and recent star formation. A description of all parameters used for the lognormal star formation history model is presented in Table~\ref{tab:bagpipes_sed_priors}.

\begin{table*}
\centering
\begin{threeparttable}
\caption{A description of each of the parameters used for the lognormal star formation history model, as well as the priors used.}
\label{tab:bagpipes_sed_priors}

\renewcommand{\arraystretch}{1.15} % a bit more vertical space
\begin{tabularx}{\textwidth}{@{} l l X l @{}}
\toprule
\textbf{Component} & \textbf{Parameter name} & \textbf{Description} & \textbf{Uniform prior ranges} \\
\midrule

\multirow{4}{*}{SFH: lognormal}
& $t_{\rm max}$ (Gyr) & Age of Universe at peak star formation & $(0.1,\,15)$ \\
& $\mathrm{FWHM}$ (Gyr) & Full width at half maximum of SFH & $(0.1,\,20)$ \\
& $\log(M_\star/M_\odot)$ & Total stellar mass formed & $(1,\,15)$ \\
& $Z_\star/Z_\odot$ & Stellar metallicity & $(0,\,3)$ \\
\midrule

\multirow{1}{*}{Nebular}
& $\log_{10}U$ & Ionization parameter & $(-4,\,-2)$ \\
\midrule

\multirow{4}{*}{Dust}
& $A_V$ (mag) & Dust attenuation coefficient & $(0,\,4)$ \\
              &  $q_{\rm PAH}$ &  The PAH mass fraction & $(0.1,\,4.58)$ \\
              & $u_{\min}$ &  The minimum starlight intensity dust is exposed to & $(0.1,\,25)$ \\
              & $\gamma$ & Fraction of stars at $u_{\min}$ & $(0.0005, 1.0)$ \\

\bottomrule
\end{tabularx}
\end{threeparttable}
\end{table*}

\subsubsection{Resolved stellar mass surface densities}
\label{resolved_sigma}

To construct the radial acceleration relation, we require the stellar mass surface density as a function of galaxy radius, $\Sigma_{*}(r)$. 
We derive this through resolved SED
fitting in concentric annuli using \textsc{Bagpipes}, employing two
independent methods to extract the input photometry. In the first, direct approach, the
fluxes are measured directly from the masked, background-subtracted
images within elliptical annuli, as described in
Section~\ref{direct_photometry}. In the second, S\'ersic based approach, for each galaxy, the annular fluxes in each band are computed by evaluating the best-fitting parameters from the model (single or preferred double S\'ersic) at the  radii at which the H{\sc i} rotation curve is sampled. 
In both cases, at each annulus, \textsc{Bagpipes} receives an SED and returns a stellar mass. The SED shape (flux ratios between bands) determines the mass-to-light ratio, $\Upsilon_\star$, while the absolute flux level sets the mass. We adopt the direct photometry results as our fiducial method, since it makes no assumptions about the radial profile shape, and present the S\'ersic based results as a comparison check in all that follows.

\begin{figure}
    \centering
    \includegraphics[width=0.5\textwidth]{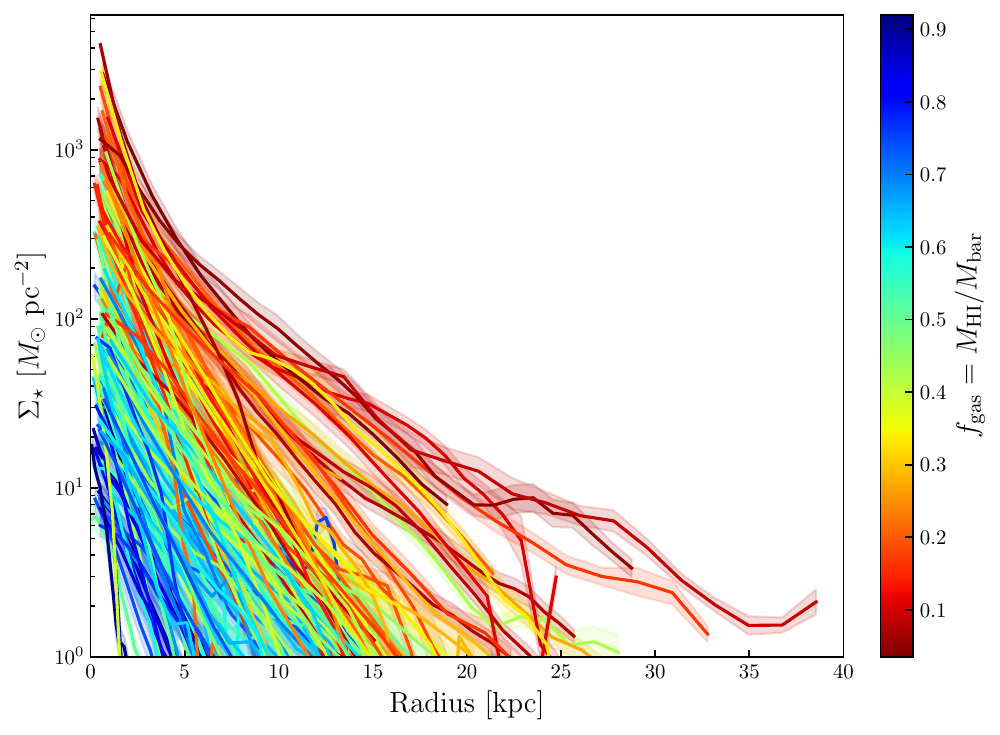}
    \caption{Resolved stellar surface mass densities for our sample of galaxies, colour-coded by their atomic gas fraction, $\mathrm{f_{g} = M_{HI}/M_{bar}}$, with shaded regions indicating uncertainties.}
    \label{fig:stellar_surface_dens}
\end{figure}

Figure~\ref{fig:stellar_surface_dens} presents all the stellar mass surface densities obtained for our sample, colour-coded by their atomic gas fraction -- high stellar mass galaxies with lower atomic gas fractions (i.e., profiles are higher and redder in the figure) tend to have shallower declines in their surface brightness profiles , indicative of stellar contribution at all radii, whereas low-mass, gas-dominated galaxies show steeper declines.

\subsubsection{Mass-to-light ratio variations}
\label{ml-variations}
The assumed stellar mass-to-light ratio ($\mathrm{\Upsilon}_{\star}$) is the dominant source of uncertainty in rotation curve decomposition, as it directly controls the inferred distribution of baryonic matter. While a spatially constant $\mathrm{\Upsilon}_{\star}$ is commonly adopted in dynamical studies, the systematic biases introduced by this assumption have been quantified in only a few works \citep{Ponomareva_2017,Liang_2024, rar_paper_2025}. Since $\mathrm{\Upsilon}_{\star}$ depends on stellar age, metallicity, dust extinction, and the assumed IMF \citep{Bruzual_Charlot_2003, Bell_2003}, and varies significantly with wavelength — particularly in bluer filters — we investigate radial variations for all galaxies in our sample rather than imposing a fixed value, following the methodology described in \cite{rar_paper_2025} and outlined briefly again here.

We compute $\mathrm{\Upsilon}_{\star}$ in the near-infrared VISTA $K_s$ band at 2.2\,$\mu$m, where the light is dominated by the older stellar populations that constitute the bulk of the stellar mass and is minimally affected by dust attenuation \citep{Into_2013, Meidt_2014, McGaugh_2015, Sorce_2013}. This band is also close in wavelength to the 3.6\,$\mu$m {\em Spitzer} band commonly used in similar studies \citep[e.g.][]{Verheijen_2001}. Although we have {\em Spitzer} data, we choose the 
VISTA $K_s$ band due the similar resolution to the visible-wavelength data, and higher resolution compared to {\em Spitzer}, allowing a more robust analysis of resolved mass-to-light ratios across our sample.  

\begin{figure}
    \centering
    \includegraphics[width = \linewidth]{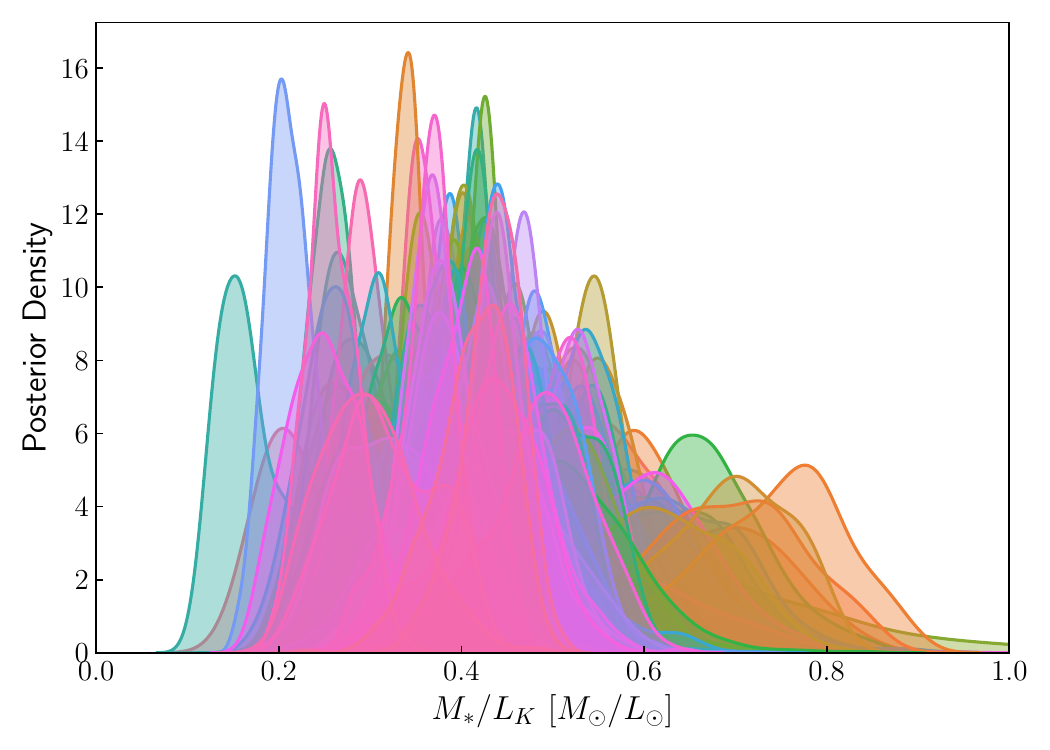}
    \caption{Distribution of mass-to-light ratios ($\mathrm{\Upsilon_{\star}}$) in the $K_s$-band from the mass-to-light ratio posterior samples derived from SED fitting with {\sc Bagpipes}. The sample has  a median of $0.43^{+0.11}_{-0.10}$.}

    \label{fig:total_ml_kde}
\end{figure}

First, as in previous work, we compute the distribution of \textit{total} $K_{s}-$band mass-to-light ratios for our sample, which is shown in Figure~\ref{fig:total_ml_kde}. Galaxies with prominent spiral arms and star-forming regions have on average lower mass-to-light ratios than early type spirals (Sa, SBa). We find values spanning from 0.29 to 0.60, with a median of 0.45 $M_{\odot}L_{\odot}^{-1}$, values consistent with \citet{rar_paper_2025}, who employed a similar SED-based methodology, and with the DiskMass Survey results of \citet{Martinsson_2013} who used a dynamical estimate of the mass-to-light ratios.

\begin{figure}
    \centering
    \includegraphics[width=\linewidth]{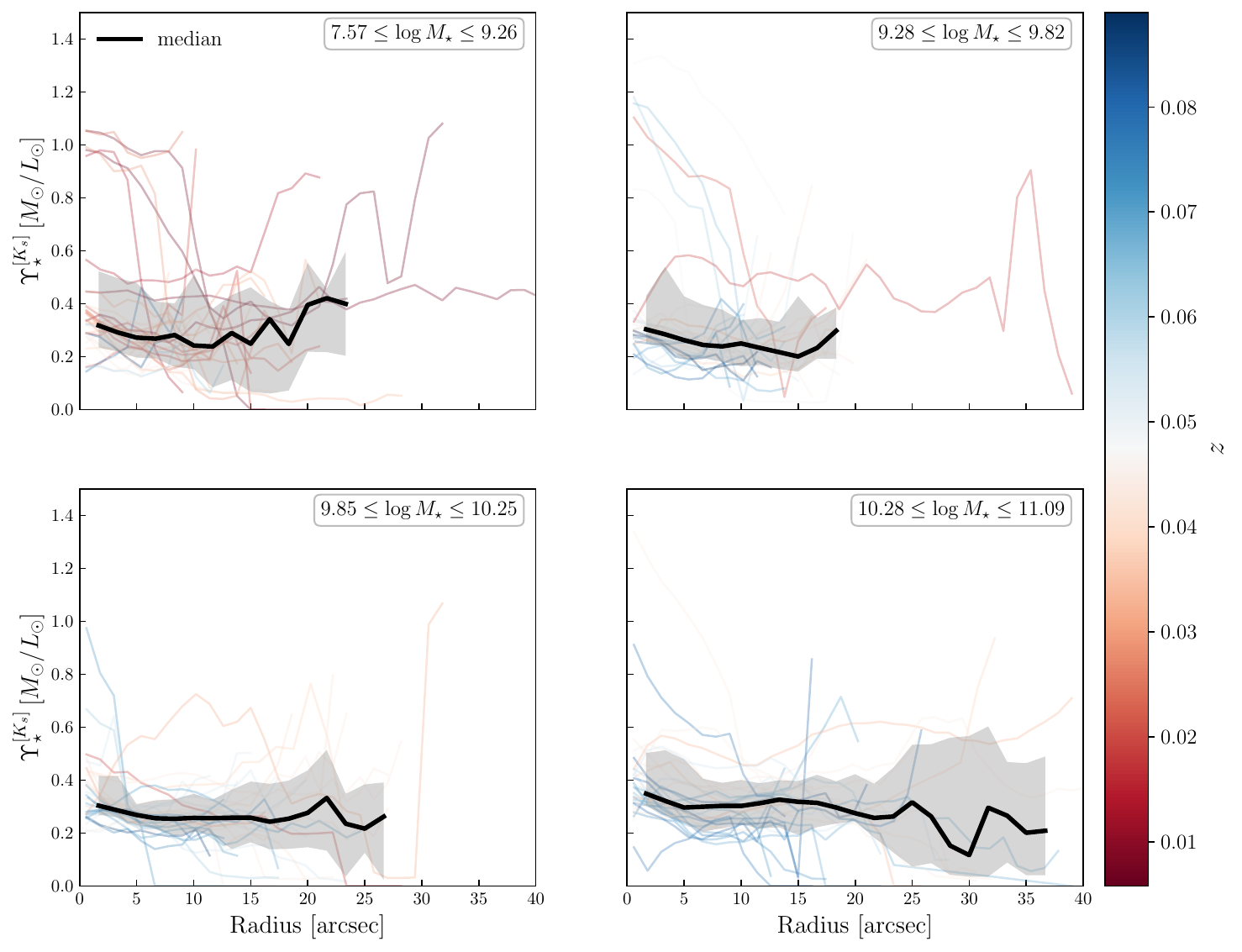}
    \caption{Mass-to-light ratio variations in $K_s$-band as a function of radius for all galaxies in our sample in 4 stellar mass bins, colour coded by their redshift. The black solid lines represent the median mass-to-light ratio relation in the corresponding stellar mass bins, with grey shaded bands showing the 16th-84th percentile spread of individual galaxies' $\mathrm{\Upsilon}_{\star}$ at each radius. Unsurprisingly, the mass-to-light ratios are typically higher in the center, followed by a decrease and flattening at large radii, where the surface brightness exponentially decreases.}
    \label{fig:ml-ratio-variations}
\end{figure}

Next, to determine the radially varying mass-to-light ratio, we take the ratio of the stellar mass, $\Sigma_{*}(r)$,  obtained from SED fitting to the $K_s$-band surface brightness, both inclination corrected,  derived from our fiducial non-parametric photometry. 
Figure~\ref{fig:ml-ratio-variations} shows the resulting radial profiles for all galaxies in our sample, colour-coded by redshift. The profiles reveal a common trend where the mass-to-light ratio is generally higher in the central regions -- reflecting the older, redder stellar populations concentrated in galaxy bulges -- and decreases outward before flattening at large radii where the disc light dominates, although we note there are large variations on a galaxy to galaxy basis.
These radial gradients in $\mathrm{\Upsilon}_{\star}$ further demonstrate that a spatially constant mass-to-light ratio is insufficient to capture the stellar population variations present in disc galaxies, and motivate our use of resolved SED fitting rather than a single global value $\mathrm{\Upsilon}_{\star}$ when converting light to stellar mass for determining the acceleration due to the baryonic mass, as in previous work. 

\subsection{H{\sc i} surface density profiles and rotation curves}
\label{HI-kinematics}
We derive rotation curves and radial H\textsc{i} surface density profiles using \textsc{3D Barolo} (3D-Based Analysis of Rotating Objects from Line Observations; \citealt{DiTeodoro_2015}), which fits three-dimensional 3D tilted-ring models directly to emission-line data cubes. \textsc{3D Barolo} can reliably recover the underlying kinematics of galaxies with as few as three resolution elements along the major axis \citep{DiTeodoro_2015, Mancera_Pina_2020}, making it well suited to our marginally resolved H\,\textsc{i} data. We refer the reader to \cite{rar_paper_2025} for a detailed description of the fitting procedure and summarise the key choices here.

The source mask is generated using the SMOOTH algorithm with a signal-to-noise threshold of 3. We adopt azimuthal normalisation (NORM=AZIM) to obtain the average gas surface density for each ring, and assume a razor-thin H\,\textsc{i} disc throughout. Since many of our H\,\textsc{i} data are only marginally resolved, we fix the inclination to optically derived values. 
These are computed from the 
G-band axis ratio using the standard relation (Equation~\ref{eq:inclination}) adopting $q_0=0$ (thin disc); we verified that alternative prescriptions for $q_0$ produce negligible differences well within the uncertainties. The fitting proceeds in two stages: an initial fit with rotation velocity, velocity dispersion, inclination, and position angle all free, followed by a second fit in which only the rotation velocity and velocity dispersion are varied while the geometric parameters are held fixed. Rings are spaced at half the beam width, yielding two points per beam. Uncertainties on the rotation velocities are estimated via \textsc{3D Barolo}'s built-in Monte Carlo method (FLAGERRORS). 
From the same fits, we extract the azimuthally averaged, inclination-corrected H\,\textsc{i} surface mass densities at each radius, which \textsc{3D Barolo} converts to physical units of
 $\mathrm{M_\odot\,pc^{-2}}$ \citep{Meyer_2017}.

\begin{figure}
    \centering
    \includegraphics[width =\linewidth]{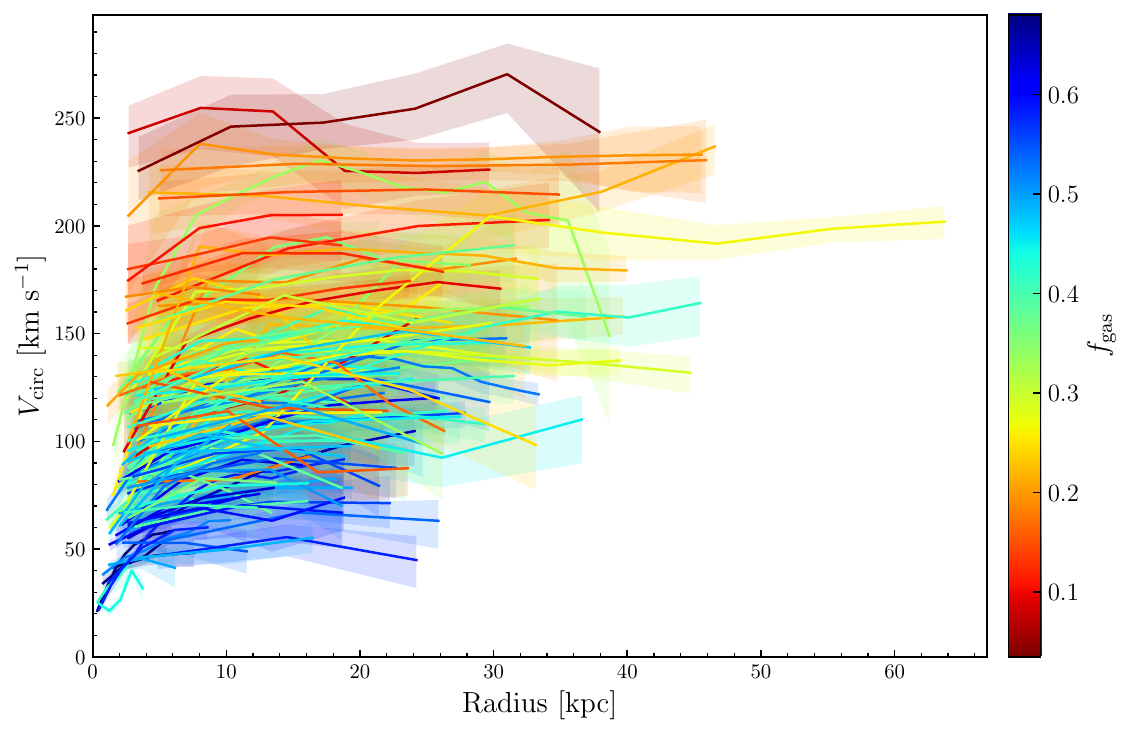}
    \caption{Rotation curves for our sample of 130 galaxies, colour coded by their gas fraction, $\mathrm{f_{g} = M_{HI}/M_{bar}}$, with measurement uncertainties indicated by the shaded regions.}
    \label{fig:rotation_curves}
\end{figure}

\begin{figure}
    \centering
    \includegraphics[width =\linewidth]{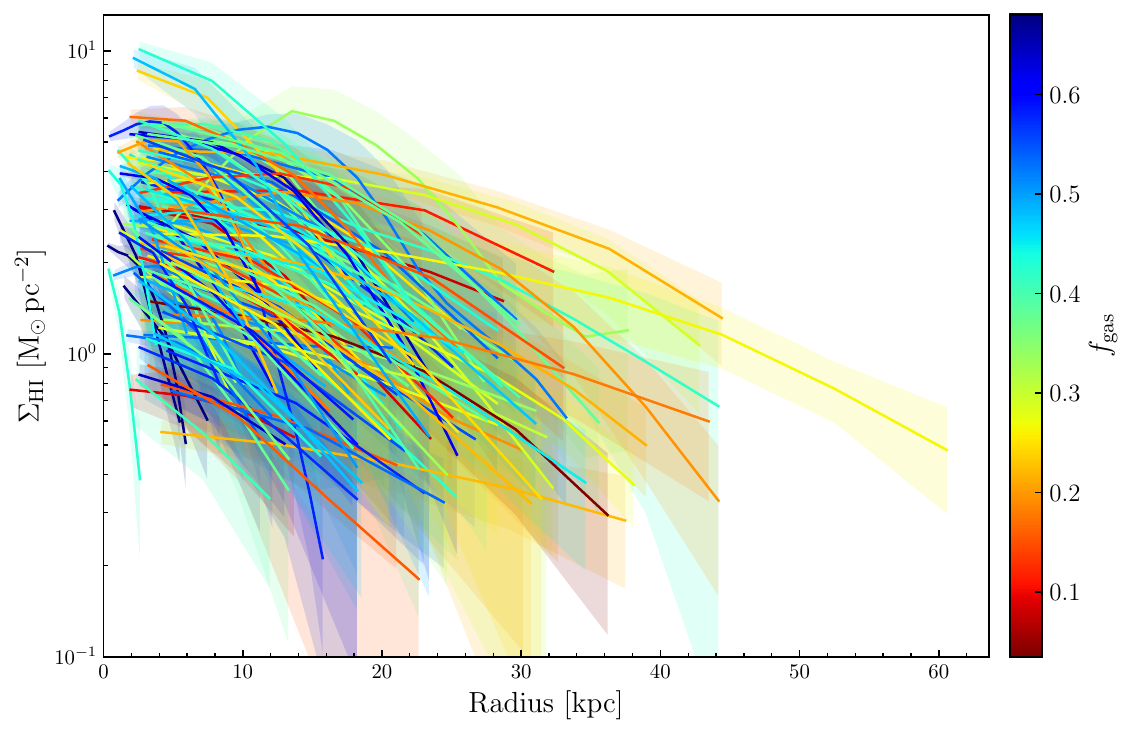}
    \caption{H{\sc i} mass surface densities for our sample of 130 galaxies, colour-coded by their gas fraction, $\mathrm{f_{g} = M_{HI}/M_{bar}}$, with measurement uncertainties indicated by the shaded regions.}
    \label{fig:hi_surf_dens}
\end{figure}

The H{\sc i} surface mass densities and rotation curves for our 130 galaxy sample are presented in Figure~\ref{fig:rotation_curves} and ~\ref{fig:hi_surf_dens}, colour-coded by their gas fraction. 

\subsection{Baryonic circular velocities}
\label{circular-velocities}
We compute the circular velocity contributions of the stellar and gas components by numerically solving Poisson's equation using {\sc 
Galpynamics}, which takes as input the mass surface density profile and returns the circular velocity at each radius \citep{Iorio_2018}, following the methodology described 
in detail in \citet{rar_paper_2025}. We summarise the key choices here.

\subsubsection{Stellar disc}
The morphologies in our sample are predominantly disc-dominated: most galaxies are best described by low S\'ersic indices ($n \lesssim 2)$, indicative of pseudo-bulges, with few showing classical ones.
We therefore model the stellar surface mass density profiles with a flexible fourth-degree poly-exponential function \citep{Bacchini_2019, Mancera_Pina_2022}.

\begin{equation}
  \Sigma_{\star}(R) = \Sigma_{0}\,\exp\!\left(-\frac{R}{R_d}\right)
  \left(1 + c_1 R + c_2 R^{2} + c_3 R^{3} + c_4 R^{4}\right),
\end{equation}
which captures deviations from a simple 
exponential, including central dips or peaks in the stellar profiles. 
We adopt an exponential vertical profile with scale height 
$z_d = 0.196\,R_d^{0.633}$ following \citet{Bershady_2010}, having 
verified that the choice of vertical profile ($\mathrm{sech}^2$ 
versus exponential) has negligible impact on the resulting RAR.

\subsubsection{Atomic gas disc}
For the gas component, we assume a razor-thin disc, justified by our use of H\,{\sc i} as a dynamical tracer and the negligible 
effect of disc flaring on the gravitational potential in this mass 
regime \citep{Mancera_Pina_2022}. The radial gas profiles 
are modelled with either a simple exponential or a first-order 
poly-exponential, depending on the complexity of the observed 
profile, selecting between them via the Bayesian Information Criterion (BIC). 

\subsubsection{Molecular gas}
\label{molecular_gas}
 Although direct molecular gas observations are not available for our
sample, we estimate the molecular gas contribution using the scaling
relations of \cite{Saintonge_2017} and \cite{Tacconi_2018}, which parametrise the molecular
gas depletion time as a function of redshift, stellar mass, and
offset from the star-forming main sequence:
\begin{align}\label{eq:tacconi}
\log t_{\rm depl} &= A + B\,\log(1+z) \nonumber\\
&\quad + C\,\log\!\left[
\frac{\mathrm{sSFR}}
{\mathrm{sSFR}_{\mathrm{MS}}(z,M_\star)}
\right]
+ D\,(\log M_\star - 10.7),
\end{align}
where $A = 0.21$, $B = -0.98$, $C = -0.49$, $D = 0.03$, and the
main-sequence specific star formation rate sSFR is computed from the parametrisation of
\citet{Speagle_2014}, adopted by \citet{Tacconi_2018}, with

\begin{equation}
\begin{split}
\log_{10}\!\left[\frac{\mathrm{sSFR}_{\mathrm{MS}}(z, M_\star)}{\mathrm{Gyr}^{-1}}\right]
={}& (-0.16 - 0.026\,t_c)\,(\log M_\star + 0.025) \\
   & - (6.51 - 0.11\,t_c) + 9,
\end{split}
\label{eq:speagle}
\end{equation}
where $t_c$ is the cosmic time (age of the Universe) at redshift $z$, in Gyr,
\begin{equation}
\begin{split}
\log\!\left(\frac{t_c}{\mathrm{Gyr}}\right)
={}& 1.143 - 1.026\,\log(1+z) \\
   & - 0.599\,\log^2(1+z) + 0.528\,\log^3(1+z),
\end{split}
\label{eq:tc}
\end{equation}
and $M_\star$ is in solar masses. The offset from the main sequence in
Equation~(\ref{eq:tacconi}) is evaluated as $\log(\mathrm{sSFR}/\mathrm{sSFR_{MS}})$
with the observed $\mathrm{sSFR} = \mathrm{SFR}/M_\star$, clipped to the
calibration range of \citet{Tacconi_2018}. 
The molecular gas mass is then $M_{\rm mol} =
t_{\rm depl} \times \mathrm{SFR}$, where both quantities ($\rm{SFR}$ and $M_\star$) are drawn
from the \textsc{Bagpipes} posteriors. We define the molecular gas
fraction $\mu_{\rm mol} = M_{\rm mol}/M_\star$ and distribute
$\Sigma_{\rm mol}(R) = \mu_{\rm mol}\,\Sigma_\star(R)$, following
the assumption that the molecular gas traces the stellar disc
\citep{Leroy_2008, Bacchini_2019}. For galaxies with $\log M_\star < 9.0$, where the
\citet{Tacconi_2018} calibration is poorly constrained, we adopt  $\mu_{\rm mol} = 0.07$ following the same prescription used in \citet{rar_paper_2025}; see also \citet{McGaugh_2020}.

The total gas circular velocity is then computed as a two-component sum:
\begin{equation}
v_{\rm gas, tot}^2 = 1.4\,v_{\rm HI}^2 + v_{\rm mol}^2,
\end{equation}
where the factor of 1.4 accounts for the helium contribution to the H\,{\sc i} disc mass \citep{Arnett_1999}. We note that there is no corresponding factor applied to $v_{\rm gas, mol}^2$ because the molecular gas mass already includes helium through the Tacconi depletion-timescale parametrisation.

\subsubsection{Baryonic velocity profile}
The total baryonic circular velocity is obtained by summing in quadrature the contributions of each mass component,
\begin{equation}
    v_{\rm bar}^2(r) = v_\star^2(r) + 1.4\,v_{\rm HI}^2(r) + v_{\rm mol}^2(r).
    \label{eq:vbar_sum}
\end{equation}

Uncertainties on the baryonic circular velocities are estimated through resampling. For each of 100 realisations, we draw the surface density profile from a truncated normal distribution (enforcing $\Sigma \geq 0$) and, for the stellar disc, the scale length from its fitted uncertainty, then recompute the circular velocity using {\sc Galpynamics}. We assume a poly-exponential thick disc for the stellar component, and a razor-thin disc for the gas. The 16th-84th percentile of the resulting velocity distribution is adopted as the $1\rm \sigma$ uncertainty. 
We do not correct 
for asymmetric drift, which is negligible for the high rotational 
velocities that dominate our sample \citep{Iorio_2017, Mancera_2021}.
\citep{Dalcanton_2010} show that gas pressure support can be nonetheless significant in low mass galaxies with $v_{\rm rot} \lesssim 80\,\mathrm{km\,s^{-1}}$, a regime populated by a small fraction of our sample (Figure~\ref{fig:rotation_curves}). We verified that applying an asymmetric drift correction to these lower-mass systems leaves our results unchanged within the quoted uncertainties.

\section{Results}
\label{results}

The contributions from stars and gas are added together, and the derivative of the potential gives the baryonic acceleration:

\begin{equation}
   g_\mathrm{bar}(r) = -\frac{\partial \Phi}{\partial r} = \frac{v_\mathrm{bar}^{2}(r)}{r} =  {\frac {v_\star^2 + v_{\rm gas,tot}^2}{r}}
   \label{eq:g_bar_eq}
\end{equation}
where $g_{\rm bar}$ is the acceleration due to the baryonic mass, $v_{\rm bar}$ is the circular velocity resulting from the baryonic mass, both determined at radius $r$. 
The total centripetal acceleration, derived from rotation curves, is given by: 
\begin{equation}
    g_\mathrm{obs}(r) = \frac{v_\mathrm{rot}^2(r)}{r},
    \label{eq:g_obs_eq}
\end{equation}
where $v_{\rm rot}(r)$ is the velocity measured from the rotation curve at radius $r$.

\begin{figure*}
    \centering
    \includegraphics[width=\textwidth]{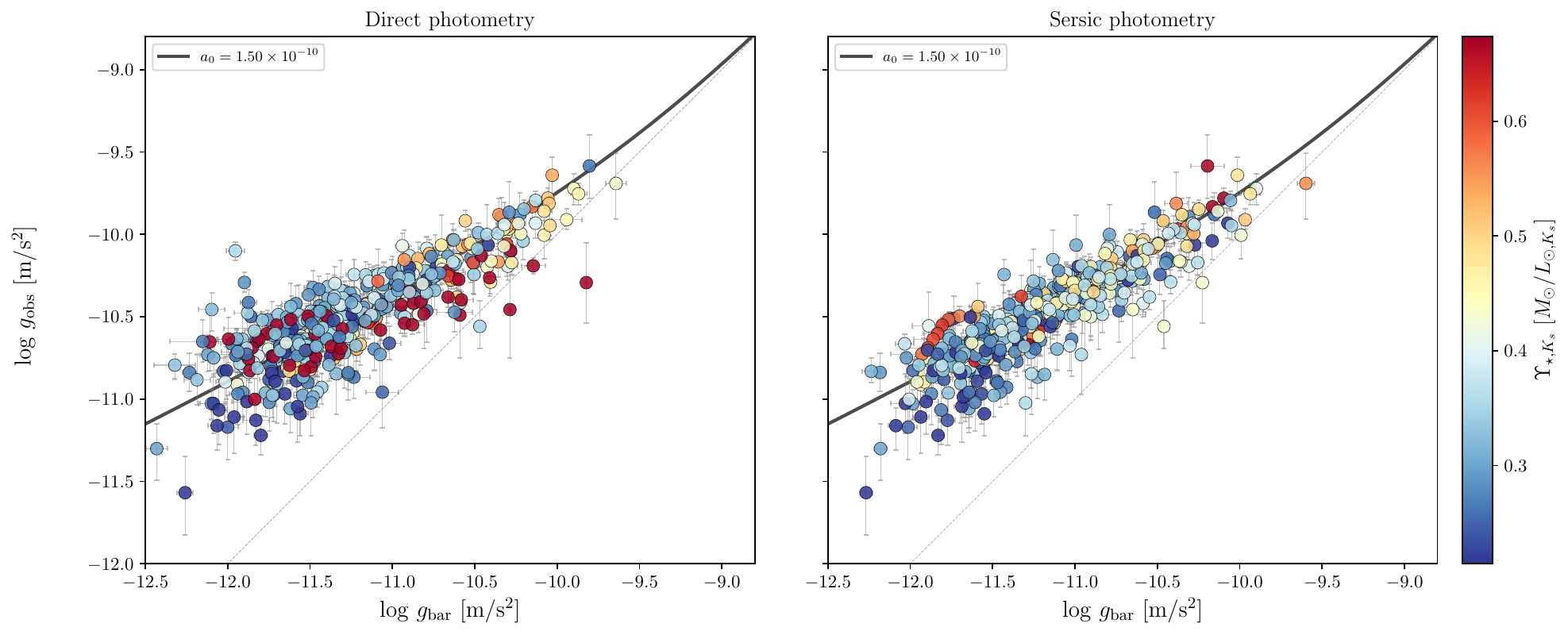}
    \caption{{\it Left}: The RAR for our sample of 130 late-type galaxies obtained
    for the direct photometry case, colour-coded by their resolved mass-to-light
    ratio in the $K_s$ band, $\Upsilon_{\star,K_s}$ [$M_\odot/L_{\odot,K_s}$].
    {\it Right}: The RAR with $g_{\rm bar}$ obtained through S\'ersic fits,
    colour-coded by their resolved mass-to-light ratios. The black solid line is the best fit RAR IF for our fiducial case; the grey dotted line represents the one-to-one line ($g_{\rm obs}$ =  $g_{\rm bar}$), where the baryons alone account for the observed dynamics.}
    \label{fig:rar_direct_sersic_comp}
\end{figure*}

The baryonic and dynamical acceleration components allow us to analyze the radial acceleration relation at all radii. 
Due to the angular resolution of our H{\sc i} data, we discard the innermost points, which correspond to less than 5 arcseconds in radius. The radial acceleration relation for our sample, derived using the varying mass-to-light ratio using both methods---S\'ersic and direct photometry---is shown in Figure~\ref{fig:rar_direct_sersic_comp}, colour coded by the corresponding mass-to-light ratio at each radius.

\subsection{Fitting the RAR}
\label{rar-fits}

We fit the RAR with the MOND-inspired relation from \cite{McGaugh_Lelli_2016} (``RAR'' or ``McGaugh--Lelli--Schombert'' interpolating function), described by the equation:

\begin{equation}
    \mathrm{g_{obs}} = \mathrm{F (g_{bar})} = \mathrm{\frac{g_{bar}}{1 - e^{-\sqrt{g_{bar}/a_{0}}}}},
    \label{eq:RAR-eq-mond}
\end{equation} where $a_{0}$ represents the acceleration scale.

The fit is performed with the Python package {\sc Roxy} \citep{roxy}, which implements the ``Marginalised Normal Regression (MNR)'' method. 
MNR fits a function to data accounting for uncertainties in both $x$ and $y$ directions, intrinsic scatter in the relation and unknown ``true'' $x$ values through the use of a Gaussian hyperprior with inferred mean $\mu_\text{gauss}$ and standard deviation $w_\text{gauss}$. We apply MNR to the accelerations in base 10 logarithmic space, so that both $\mu_\text{gauss}$ and $w_\text{gauss}$ are expressed in dex. Extensive mock tests showed that MNR, unlike most other methods employed in the literature, is unbiased~\citep{roxy}. The likelihood is sampled using the No U-Turn Sampler (NUTS) method of Hamiltonian Monte Carlo. 
We adopt uniform priors for $\log_{10}(a_0)$ between $-15$ and $5$, and for the intrinsic scatter between $0$ and $3$ dex. 
We present results for two photometric pipelines: our fiducial 
non-parametric direct-aperture photometry (Section~\ref{direct_photometry}), 
and the parametric S\'ersic fits (Section~\ref{photometry}), to assess the sensitivity of the RAR to our baryonic modelling. 

We run {\sc Roxy} with 700 warm-up steps and 5000 samples, and check that in all cases this produces a Gelman--Rubin statistic $\hat{R} \le 1.01$. For our fiducial direct photometry analysis, we recover a best-fit value of the acceleration scale $a_0 = (1.50 \pm 0.05) \times 10^{-10}\ \mathrm{m\,s^{-2}}$ and an intrinsic scatter of $0.096 \pm 0.006$ dex for the full sample.
For the S\'ersic based fit, we obtain $a_0 = (1.48 \pm 0.04) \times 10^{-10}\ \mathrm{m\,s^{-2}}$ and an intrinsic scatter of $0.075 \pm 0.006$. The acceleration scales are in excellent agreement, at the $0.1\sigma$ level, demonstrating that $a_0$ is robust to the choice of photometric prescription; the intrinsic scatter however, is mildly sensitive to it at approximately $2\sigma$ level, with the smoother S\'ersic profiles producing lower scatter, most likely by suppressing some real galaxy structure and radial variation that the aperture photometry preserves (see Section~\ref{rar_systematics} below for a more detailed discussion).

For comparison, the COSMOS sub-sample of 19 galaxies analysed in \citet{rar_paper_2025} yielded an acceleration scale $a_0 = (1.69 \pm 0.13 \times 10^{-10} ~\mathrm{m\, s^{-2}}$ and $\sigma_{\rm int} = 0.045 \pm 0.022$ dex. Our expanded sample prefers a slightly lower $a_0$, consistent with our previous result within $\sim 1\sigma$.
Our intrinsic scatter is larger than $0.045 \pm 0.022$~dex of the 19-galaxy COSMOS sample and is now clearly
non-zero statistically. This value remains consistent with recent estimates from SPARC galaxies \citep{Lelli_2017, Chae_2021, Chae_2022,  Desmond_2023}, supporting the conclusion that the RAR is a tight and potentially fundamental scaling relation \citep{Lelli_2017, Desmond_2023, Stiskalek_2023}. 

\subsection{Systematics}
\label{rar_systematics}
% given the 
We now examine the sensitivity of the RAR to the chosen baryonic mass methodology: the photometry pipeline used to measure stellar surface mass density profiles (and hence the treatment of the radial mass-to-light ratio), and the inclusion of a molecular gas component in the baryonic budget through the use of scaling relations. 

Potential systematics in our analysis reside in the construction of the stellar mass surface density profile, $\Sigma_\star(r)$, which we measure 
with two independent pipelines. The S\'ersic-based photometry used in SED fitting imposes a smooth profile, washing out structure in more complex galaxies with irregular morphologies, or other non-axisymmetric features such as bars etc, whilst the direct aperture photometry in elliptical annuli preserves it, at the cost of being more sensitive to noise of faint regions in galaxy outskirts. 

\begin{figure*}
    \centering
    \includegraphics[width=\textwidth]{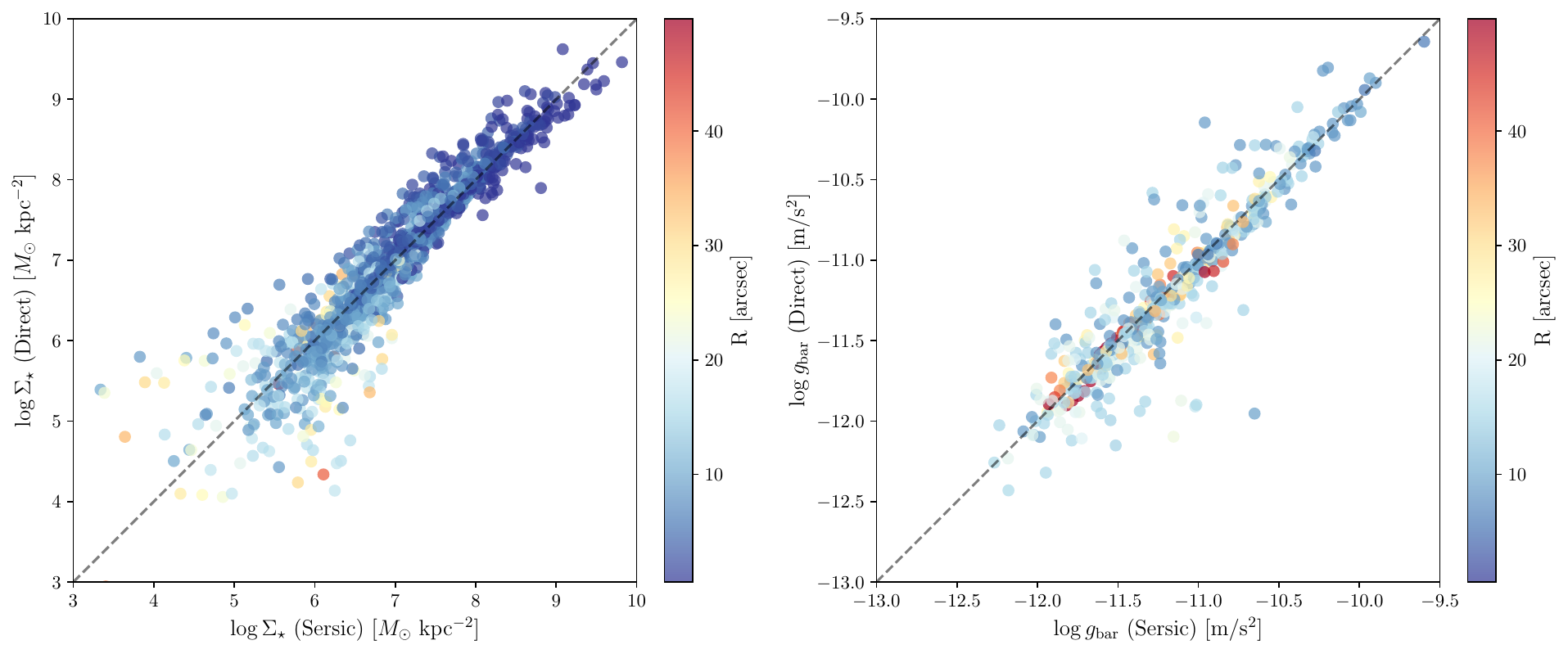}
    \caption{{\it Left}: $\rm \Sigma_{star}$ comparison across all photometric annuli from the S\'ersic fitting and direct photometry pipelines. {\it Right}: $\rm g_{bar}$ comparison at all kinematically sampled radii. Both quantities are colour-coded by radius. The dashed black lines in both panels correspond to the 1:1 line.}
    \label{fig:sigma_bar_gbar_comp}
\end{figure*}

Figure~\ref{fig:sigma_bar_gbar_comp} compares the result of the two pipelines at every photometric annulus and every kinematic radius for the full sample.

The left panel shows the stellar mass surface density profiles $\Sigma_\star$ from the S\'ersic fit against 
the corresponding direct-aperture measurement, while the right panel 
shows the resulting $g_{\rm bar}$ at each rotation-curve radius. Both quantities are colour-coded by their galactocentric radius in arcseconds. 

The data points cover the inner, mid and outer disc, and they show excellent agreement in the inner parts, where they cluster tightly along the 1:1 line (blue data points), with discrepancies arising in the mid and mostly outer annuli. 
The two pipelines diverge most strongly in the outer annuli (red data points), which is expected since the S\'ersic profile is less constrained by the data and extrapolates the model outwards, whilst the direct photometry is most affected by background noise, incomplete contaminant masking and low surface brightness regions. 

A key distinction here is that $\Sigma_\star (r)$ is a per-annulus measurement, whereas $g_{\rm bar}$ on the other hand is not: for an axisymmetric disc the baryonic acceleration at radius $r$
depends on the mass distribution at all radii and the gradient of $\Sigma_\star (r)$ \citep{Casertano_1983}. For an exponentially declining stellar profile, $\Sigma_\star(r)$ is faintest at the largest radii, therefore the outermost radii contribute little to the integrated $g_{\rm bar}$. This is why the two photometric pipelines can disagree substantially in $\Sigma_\star(r)$ at large radius while still yielding consistent values of $g_{\rm bar}$.
For galaxies with low stellar masses, the stellar contribution to the baryonic acceleration has little effect, with $g_{\rm bar}$ instead dominated by the gas mass, derived directly from the H{\sc i}, not photometry. This results in the general agreement between the two baryonic accelerations seen in the right panel of Fig~\ref{fig:sigma_bar_gbar_comp}.
To further quantify how this radial-mass disagreement propagates into the RAR, we classify galaxies according to their per annulus consistency between the two methods, measured relative to their uncertainties. 
For each galaxy we compute the reduced chi-squared of the difference between the S\'ersic and direct $\Sigma_{\rm \star}$ profiles, as $\chi^2_{\rm red} = \sum[(\Sigma_{\rm d} - \Sigma_{\rm s})/\sigma]^2/N$, where $\sigma$ combines the per-annulus {\sc Bagpipes} posterior uncertainties of the two methods, which does not take into account any systematic uncertainties from choice of SFH or IMF, and $N$ is the number of matched annuli. 
Galaxies with $\chi^2_{\rm red} \lesssim 3$ are labeled as $\Sigma_{\rm \star}$ consistent, ``well-behaved'' (61 systems) and those above as discrepant,``less well-behaved'' (69 systems). We then refit the RAR for each subsample separately, with the direct photometry as the fiducial input. The results are reported in Table~\ref{tab:rar_subsamples}. 

The well-behaved galaxies, for which the two methods agree well, yield an acceleration scale of of $a_0 = (1.80 \pm 0.08) \times 10^{-10}~\mathrm{m\,s^{-2}}$, while the unruly subsample favours a substantially lower $a_0 = (1.31 \pm 0.06) \times 10^{-10}~\mathrm{m\,s^{-2}}$ - $\sim 5\sigma$ difference. The intrinsic scatters, by contrast, are  consistent at $\lesssim 2\sigma$ between the two subsamples --$0.078 \pm 0.009$ versus $0.098 \pm 0.008$~dex.
This pattern shows that even when the photometric pipelines disagree about a galaxy, they do so in a coherent way such that the effect is a shift in the overall normalisation of the relation rather than an increase in the scatter about it.
These results suggest that the tightness of the RAR is thus relatively robust to the details of the baryonic modelling, even when the $a_0$ is not.

\begin{table*}
    \centering
    \begin{tabular}{lccccc}
    \hline
    (Sub)sample   & Method & $N_{\rm gal}$ & $a_0$ [$10^{-10}$ m s$^{-2}$] &$\sigma_{\rm int}$ [dex] \\
    \hline
    All & Direct & 130 &  $1.50 \pm 0.05$ & $0.096 \pm 0.006$ \\
    All &  S\'ersic & 130 & $1.48 \pm 0.04$ & $0.075 \pm 0.006$ \\
    Well-behaved  & Direct & 61 & $1.80 \pm 0.08$ & $0.078 \pm 0.009$ \\
    Less well-behaved & Direct & 69 & $1.31 \pm 0.06$ & $0.098 \pm 0.008$ \\
    COSMOS reference  & S\'ersic & 19 & $1.69 \pm 0.13$ & $0.045 \pm 0.002$ \\
    \hline
    \end{tabular}
    \caption{Best-fit RAR parameters for the combined MIGHTEE+LADUMA sample, split by photometric consistency classification. Galaxies are classified as ``well-behaved'' if the two methods, S\'ersic fitting and direct photometry, agree with each other within their uncertainties (Section ~\ref{photometry}), and "less well-behaved" otherwise.}
    \label{tab:rar_subsamples}
\end{table*}

\subsection{Shape of MOND interpolating function} 
 \label{RAR_IF}
  %The RAR is the empirical fingerprint of what the 
 The MOND-inspired RAR of Equation~\ref{eq:RAR-eq-mond} adopts a single 
fixed shape for the transition between the Newtonian regime ($g_{\rm bar} \gg a_0$) and the deep-MOND regime ($g_{\rm bar} \ll a_0$). 
A more general family of interpolating functions is the ``$\delta$-family'' 
of \citet{Famaey_McGaugh_2012},
\begin{equation}
    g_{\rm obs} = g_{\rm bar} \left[ 1 -  e^{-(g_{\rm bar}/a_0)^{\delta/2}} \right]^{-1/\delta},
    \label{eq:delta_family}
\end{equation} which introduces an additional shape parameter $\delta$ controlling the transition between the two regimes. The canonical RAR is recovered for $\delta = 1$, with higher $\delta$ values corresponding to steeper transitions.

This shape parameter is of particular interest because it governs MOND effects in the Solar System. MOND predicts the External Field Eﬀect (EFE), which is a dependence of a system's internal dynamics on the external field in which it is embedded \citep{Bekenstein_1984}, due to the MOND modification being a function of the \emph{total} gravitational field. In the Solar System, this produces a quadrupole correction to the Newtonian potential of the Sun, aligned with the direction to the Galactic centre \citep{Milgrom_2009_mond_effects}, whose magnitude is described by a parameter $Q_2$.
Measurements from Cassini tracking of Saturn found $Q_2 = (1.6 \pm 1.8) \times 10^{-27}$\,s$^{-2}$ \citep{Park_2026, Hees_2014, Harry_SS_quadrupole}, consistent with the Newtonian expectation of zero but not the naive MOND expectation of $\sim$30. Smaller $Q_2$ values in MOND require steeper transition of the interpolating function between the Newtonian and deep-MOND regimes, corresponding to higher values of $\delta$. 

This measurement---along with the null detection of a MONDian signal in the Wide Binary Test with Gaia data \citep{Banik_WBT}---requires $\delta \gtrsim 2$--$3$, which renders the Solar System Newtonian. By contrast, the SPARC RAR prefers $\delta \approx 1$, in $\sim 15\sigma$ tension with the quadrupole and wide binary measurements
\citep{Harry_SS_quadrupole, Park_2026}. 

\begin{figure}
    \centering
    \includegraphics[width=0.45\textwidth]{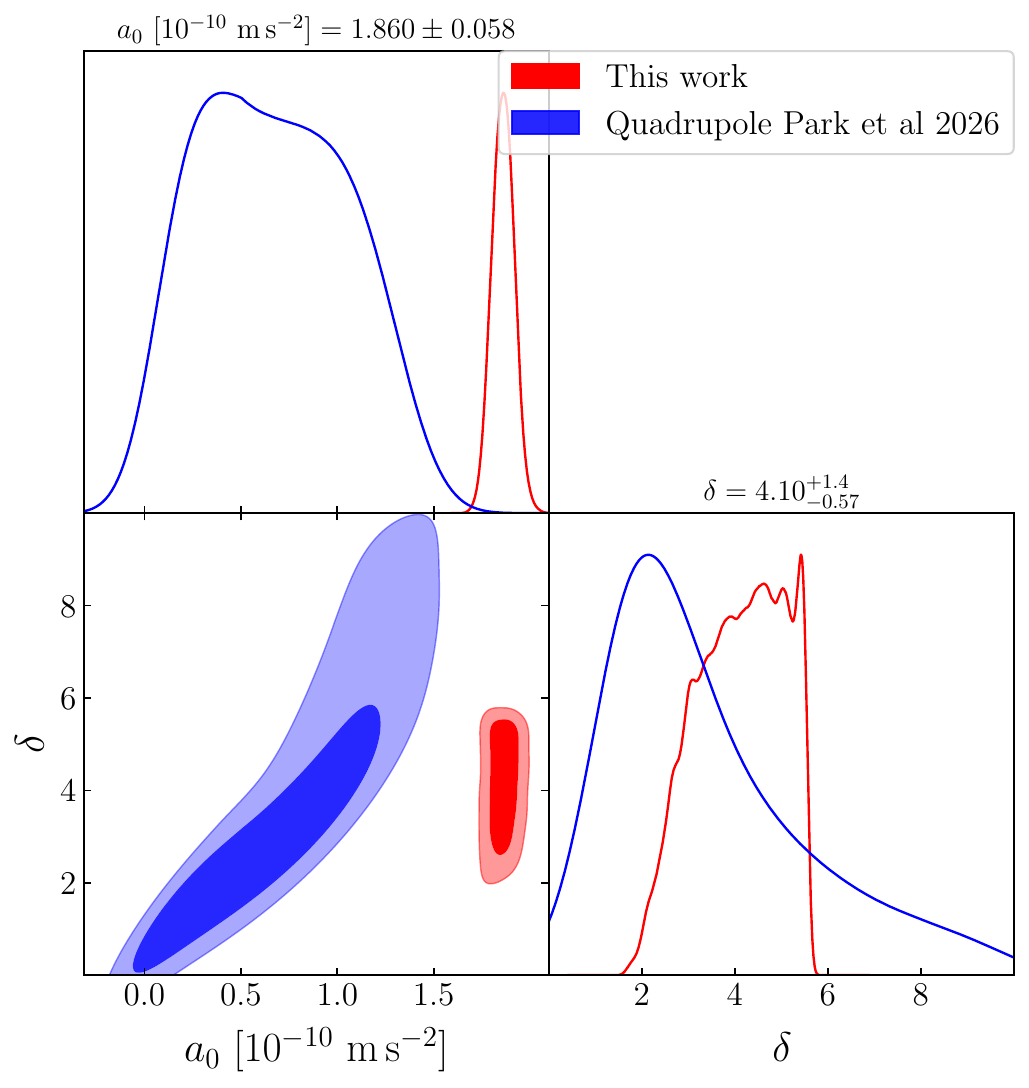}
    \caption{Posterior distribution of the acceleration scale $a_0$ and 
    shape parameter $\delta$ from the fit of Equation~\ref{eq:delta_family} 
    to our sample, and the Solar System quadrupole updated constraints~\citep{Park_2026}. 
    The 2D contours correspond to 68\% and 95\% confidence intervals.}
    \label{fig:rar_delta_corner}
\end{figure}

To investigate this in our sample, we fit Equation~\ref{eq:delta_family} using {\sc Roxy}, with uniform priors on $a_0/(10^{-10}~\mathrm{m\,s^{-2}}) \in 
[0.001, 10]$ and $\delta \in [0, 30]$. We recover 
$a_0 = (1.86 \pm 0.06) \times 10^{-10}~\mathrm{m\,s^{-2}}$ and 
$\delta = 4.1^{+1.4}_{-0.6}$, with an intrinsic scatter of $\sigma_{\rm int} = 0.111\pm 0.006$~dex. These values are statistically consistent with our previous result on the 19 galaxies COSMOS sub-sample ($a_0 = 2.00 \pm 0.15$, $\delta = 3.9^{+1.9}_{-1.1} $, 
$\sigma_{\rm int} = 0.043 \pm 0.021$~dex; \citealt{rar_paper_2025}), % 0.0949 +0.0073 -0.0068 
with the larger sample tightening the constraint on $\delta$ by a 
factor of $\sim 1.4$. The posterior distributions of $a_0$ and $\delta$ are shown in Figure~\ref{fig:rar_delta_corner}.
Our preferred $\delta \approx 4.10$ is significantly larger than the SPARC value, and is consistent with both the Cassini quadrupole measurement and the Wide Binary Test, ameliorating the 
$15\sigma$ tension reported for the SPARC sample alone. In the MOND interpretation, this would suggest
that the SPARC analysis is affected by systematics, as previously discussed in \citet{rar_paper_2025}. 

\subsection{Baryonic Tully-Fisher} 
\label{btfr}
A closely related scaling relation is the baryonic Tully-Fisher relation (bTFR), which connects the total baryonic mass of a galaxy to its asymptotic rotation velocity \citep{McGaugh_2000, McGaugh_2012, Ponomareva_2017, Lelli_2016b, Ponomareva_2021}. 
The bTFR can be understood as the asymptotic large radius or low acceleration limit of the RAR: as $R \to \infty$, $g_{\rm bar}$ is small and Equation~\ref{eq:RAR-eq-mond} gives $g_{\rm obs} \simeq \sqrt{g_{\rm bar} a_0}$. Since $g_{\rm obs}$ is related to the flat rotation velocity through: $g_{\rm obs} = V_f^2 / R$ and the baryonic acceleration is determined by the total enclosed baryonic mass $g_{\rm bar} \simeq G M_{\rm bar} / R^2$,
the radial dependence cancels and we obtain
\begin{equation}
    V_f^4 \propto G M_{\rm bar} a_0.
    \label{eq:btfr_mond}
\end{equation}

We compute the total baryonic mass as $M_{\rm bar} = M_\star + 1.4\,M_{\rm HI} + M_{\rm H_2}$, where the factor 1.4 accounts for the helium and metal contribution to the 
H\,\textsc{i} disc \citep{McGaugh_2012}, and $M_{\rm H_2}$ is 
estimated from the \citet{Tacconi_2018} scaling relations (see Section~\ref{molecular_gas}), which already incorporates a helium correction for the molecular component.
The bTFR is usually measured using an average velocity over the flat part of the rotation curve, $V_{\rm flat}$. However, our galaxies span a wide range of radial extents, from 3 beams for higher redshift galaxies, to more than 10 beams across for the closest systems. Given that the definition of flat rotation velocity is ambiguous for 3 data points \citep{Lelli_2016b, Ponomareva_2021}, we therefore adopt $V_{\rm out}$ as the rotational velocity measured at the outermost H{\sc i} rotation curve radius as a homogeneous velocity measurement across the full sample \citep{Papastergis_2016}.
Previous studies have shown that, if selection acts primarily on a mass-related quantity, the bTFR slope is affected by selection bias if a `direct' ( or `forward') fit is used---in which velocity is considered the independent variable and baryonic mass the dependent---but not if it is fit in the inverse direction \citep{Strauss_Willick, Sorce_2013, Bradford_2016}. This is directly relevant here, as our sample is H{\sc i} flux limited, with selection acting on $M_{\rm HI}$ and hence on  $M_{\rm bar}$. 
We therefore adopt the inverse fit conditioning on $M_{\rm bar}$, ($V_{\rm out}\,|\,M_{\rm bar}$) as our fiducial one, and report the forward fit alongside.

 We fit the bTFR with {\sc Roxy} using the MNR method, applied in $\log_{10}$-space. The fiducial inverse fit is:
\begin{equation}
    \log_{10}\!\left(\frac{V_{\rm out}}{\mathrm{km\,s^{-1}}}\right)
    = s' \log_{10}\!\left(\frac{M_{\rm bar}}{M_\odot}\right)+ I.
    \label{eq:btfr_inverse}
\end{equation}
We adopt uniform priors on the slope $s' \in [1/6, 1/2]$, intercept 
$I \in [-10, 10]$, and intrinsic scatter $\sigma_{\rm int} \in [0, 1]$~dex. 

\begin{figure}
    \centering
    \includegraphics[width=0.5\textwidth]{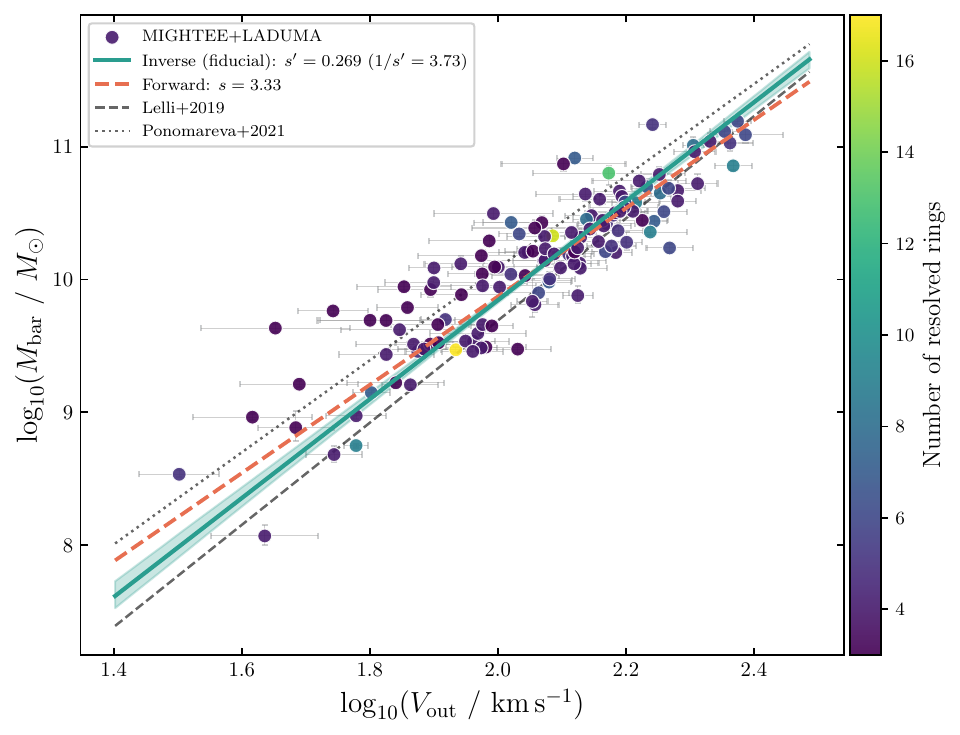}
    \caption{Baryonic Tully Fisher relation for our sample of 130 galaxies extending up to $z \sim 0.09$, colour-coded by their number of resolution elements. The dashed lines overlaid on top show the bTFR at $z = 0$ for SPARC galaxies \citep{Lelli_20P19} and the bTFR for a subsample of MIGHTEE galaxies from \citet{Ponomareva_2021}.}
    \label{fig:btfr}
\end{figure}

The baryonic Tully Fisher relation for our sample of galaxies is shown in Figure~\ref{fig:btfr} and the best-fit parameters are reported in Table~\ref{tab:btfr_fit}. The fiducial fits gives a slope $s' = 0.27 \pm 0.01$ 
and an intrinsic (vertical) scatter of  $\sigma_v = 0.054 \pm 0.011$~dex.

For a comparison with the literature, which conventionally reports the bTFR with $M_{\rm bar}$ as the dependent variable, we convert to direct-equivalent quantities: a slope $1/s' = 3.73 \pm 0.15$, a zero-point $I = 2.39 \pm 0.33$
(at $\log_{10} V_{\rm out} = 0$), and an orthogonal scatter
$\sigma_\perp = 0.053 \pm 0.006$~dex.
These are in excellent agreement with the values reported by \citet{Ponomareva_2021} based on a subsample of 67 MIGHTEE galaxies at similar redshifts using the same outer rotation velocity $V_{\rm out}$ ($s = 3.47^{+0.37}_{-0.30}$, $I = 2.10^{+0.71}_{-0.86}$), and with those found by \citet{Jarvis_2025} for H{\sc i} galaxies at $z > 0.25$, noting that these studies adopt an orthogonal fit.
Our zero-point lies $0.4$~dex ($\approx 1\sigma$) above the
\citet{Lelli_2016b} $V_{\rm flat}$ value ($1.99 \pm 0.18$); the molecular gas
contributes only $0.01$~dex of this, and we attribute the remainder to the H{\sc i} selection discussed below -- at a fixed $V_{\rm out}$, $\log M_{\rm bar}$ is slightly higher for our sample as we detect the most massive H{\sc i} sources at these higher redshifts.
Our orthogonal scatter is larger than the
$0.026 \pm 0.007$~dex of \citet{Lelli_2016b} for local SPARC galaxies; the SPARC measurements however use $V_{\rm flat}$.

As a diagnostic, we also perform the direct fit ($M_{\rm bar}\,|\,V_{\rm out}$),
which returns a shallower slope $s = 3.33 \pm 0.15$. We stress however that this value is not directly comparable to the forward-equivalent slope implied by the inverse fit: the two differ precisely because of the selection bias the inverse fit is expected to remove, and the 1.9$\sigma$ difference between them is a first indication of selection bias, quantified in Section~\ref{hi_selection}.

\begin{table}
    \centering
    \begin{tabular}{lcccc}
    \hline
     Fit & Intercept & Slope & $\sigma_y$ [dex] & $\sigma_\perp$ [dex] \\
     \hline
     Inverse  & $ -0.65 \pm  0.11 $ & $0.27\pm  0.01$ & $0.054 \pm  0.006$ & $0.053 \pm 0.006$ \\
     \hline  
     % Forward (direct) & $3.22 \pm 0.30$ & $3.33 \pm 0.14$ & $0.19 \pm  0.02$ & $0.056 \pm 0.006$ \\
     % \hline
    \end{tabular}
    \caption{Best-fit bTFR parameters from the {\sc Roxy} MNR  inverse fits---intercept, slope and vertical scatter. Note that the values are given for $V_{\rm out}$ used as the velocity for the 
    full sample of 130 galaxies.}
    \label{tab:btfr_fit}
\end{table}

\begin{figure}
    \centering
    \includegraphics[width=0.5\textwidth]{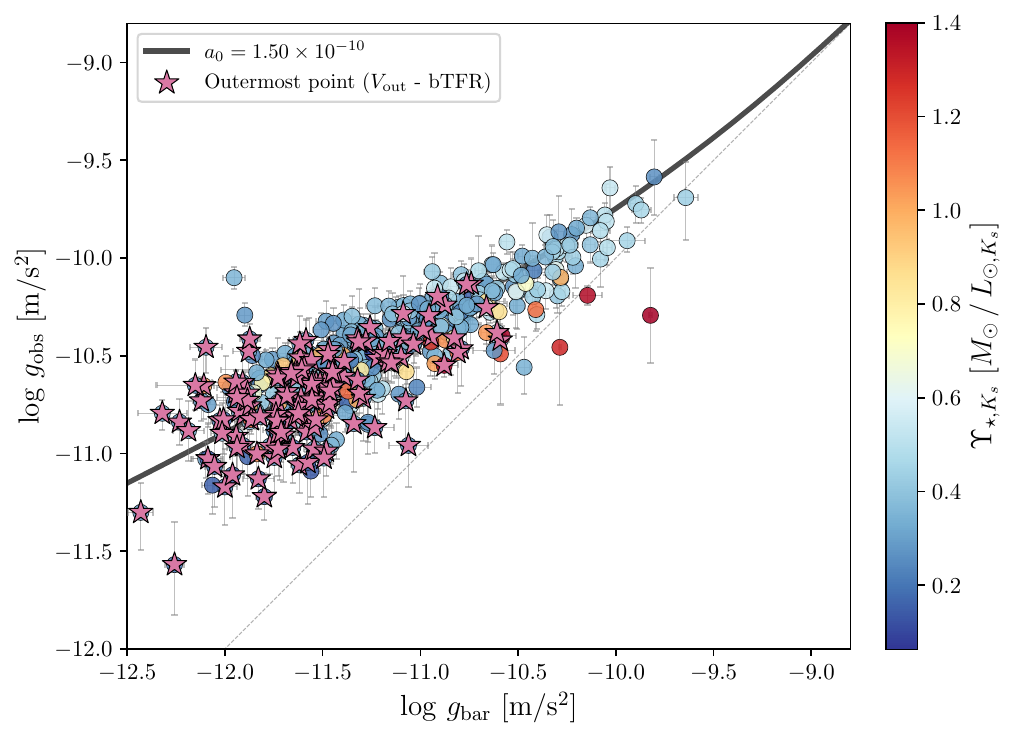}
    \caption{The RAR for our sample of galaxies extending up to $z \sim 0.09$, colour-coded by their resolved mass-to-light ratio, with the outermost points (those used in the bTFR) shown as pink stars.}
    \label{fig:rar_with_outermost_point}
\end{figure}

\subsection{Redshift evolution}
\label{z_evolution}
In this section, we investigate whether there is any significant evidence for redshift evolution in either scaling relations, the RAR or the bTFR. 
Several distinct explanations could produce an apparent redshift trend. It could reflect genuine evolution, in which case the trend should imprint consistently on both relations, irrespective of which variable is treated as the dependent one in the bTFR fit. Alternatively, it could arise from subtle systematics in how we model the baryonic mass -- for example, the stellar mass-to-light ratio prescription, the assumed interpolating function shape (Section~\ref{RAR_IF}), or simply from the H{\sc i} selection of our flux-limited sample. We first quantify the evolution itself, using both a MOND-motivated parametrisation linking the RAR and bTFR zero-points directly and a MOND-independent power-law zero-point evolution, before examining these explanations in turn in Section~\ref{explaining_z_trends}.

The methodology for the redshift-evolution fit follows that introduced in \citet{rar_paper_2025}, where it was first applied to the COSMOS sub-sample of 19 galaxies combined with SPARC to  anchor the $z \approx 0$ end, due to the relatively small sample from MIGHTEE-HI DR1. Here we are able to extend the analysis to the full sample of 130 homogeneously H{\sc i} selected galaxies spanning $0 < z \lesssim 0.09$,  and fit the evolution for our MIGHTEE and LADUMA galaxies alone. However, we also present the results after combining them with  the SPARC sample \citep{Lelli_2016}, which anchors the $z \approx 0$ end of both relations.

For the RAR, we introduce a redshift-dependent acceleration scale of the form:
\begin{equation}
    a(z) = a_0 + a_1 \times z,
    \label{eq:a_z}
\end{equation}
where $a_1$ quantifies the evolution.
For our MIGHTEE and LADUMA sample alone we find $a_0 = (1.54 \pm 0.11)$ and $a_1 = (-1.60 \pm 2.33) \times10^{-10}~\mathrm{m\,s^{-2}}$---consistent with no evolution of the acceleration scale.

Adding SPARC as a $z \approx 0$ anchor, the combined RAR fit returns $a_0 = (1.15 \pm 0.02) \times 10^{-10}~\mathrm{m\,s^{-2}}$ and $a_1 = (5.23 \pm 1.05) \times 10^{-10}~
\mathrm{m\,s^{-2}}$, a formal $5.0\sigma$ preference for an acceleration scale that increases with redshift.
This positive RAR evolution is consistent with our previous result based on a smaller sample of 19 galaxies \citet{rar_paper_2025}, and within $1.5\sigma$ of the evolution reported for the sample of 79 MUSE
star-forming galaxies at intermediate redshifts by \cite{Ciocan_2026}.
Our MIGHTEE+LADUMA only fit differs from the fit using our sample with SPARC as the $z=0$ anchor at the 2.5$\sigma$ level, suggesting that the previous detected evolution was largely driven by the different sample selection of the low redshift anchor of SPARC and the higher redshift sample H{\sc i}-selected sample from MIGHTEE.

Since $a_0$ sets the normalisation of the bTFR in MOND, any evolution of $a_0$ with redshift should be reflected in the bTFR zero-point. In the low acceleration limit, $V^4 = G\,M_{\rm bar}\,a_0$, thus

\begin{equation}
    \log_{10}\!\left(\frac{M_{\rm bar}}{M_\odot}\right) = 
    4\,\log_{10}\!\left(\frac{V}{\mathrm{km\,s^{-1}}}\right) 
    + ZP(a_0),
    \label{eq:btfr_mond_zp}
\end{equation}
% - \log_{10}(Ga_0)
where $ZP(a_0) = -\log_{10} (Ga_0)$ the zero-point, which depends on $a_0$. If $a_0$ evolves with redshift according to Equation~(\ref{eq:a_z}), this zero-point then shifts by
\begin{equation}
\begin{aligned}
    \Delta ZP(z) 
    &\equiv ZP(a(z)) - ZP(a_0) \\
    &= -\left[\log_{10} a(z) - \log_{10} a_0\right] \\
    &= -\left[\log_{10}(a_0 + a_1 z) - \log_{10} a_0\right],
\end{aligned}
\label{eq:zp_shift}
\end{equation}
which vanishes at $z=0$ and becomes negative for $a_1 > 0$ — i.e.\ a growing acceleration scale predicts a \emph{decreasing} bTFR zero-point at higher redshift. 

Motivated by this, we fit a MOND-inspired bTFR model in which the zero-point evolves through the same $a_1$ parameter as the RAR. We do not fit the absolute zero-point $ZP(a_0)$ directly, since it is degenerate with the slope; instead we pivot at the median baryonic mass of our data. Following our fiducial convention of conditioning on $M_{\rm bar}$ (Section~\ref{btfr}), we fit: 
\begin{equation}
    \log_{10}\!\left(\frac{V_{\rm out}}{V_0}\right) =
    s' \left[\log_{10}\!\left(\frac{M_{\rm bar}}{M_{\rm med}}\right)
    + \log_{10}\!\left(\frac{a_0 + a_1 z}{a_0}\right)\right],
    \label{eq:btfr_z_mond}
\end{equation}

where $s$ is the slope, $M_{\rm med} = 10^{10.19}\,{\rm M}_\odot$ is the median baryonic mass of the sample, $V_0$ is the velocity zero-point ($V_{\rm out}$ at $M_{\rm med}$ and $z=0$), and $a_0$ is fixed to the value from the corresponding RAR redshift evolution fit ($1.54$ and $1.15 \times 10^{-10}~\mathrm{m\,s^{-2}}$ for MIGHTEE+LADUMA and the combined sample respectively). 
We fit for the slope $s$, $V_0$, and $a_1$, adopting uniform priors 
$s \in [1/6, 1/2]$,  $\log_{10}(V_0/\mathrm{km\,s^{-1}}) \in [1.5, 2.5]$, $a_1 \in [-30, 30]\times 
10^{-10}~\mathrm{m\,s^{-2}}$, and intrinsic scatter $\sigma_{\rm int} \in [0, 1]$~dex. 

%combined MIGHTEE$+$LADUMA+SPARC RAR 
For our MIGHTEE+LADUMA sample alone, we find  $s' = 0.29 \pm 0.01$, $\log_{10}(V_0/\mathrm{km\,s^{-1}}) = 2.13 \pm 0.02$, and $a_1 = (-8.1 \pm 2.3)\times 10^{-10}~\mathrm{m\,s^{-2}}$: a $3.4\sigma$ preference for an acceleration scale that \emph{decreases} with redshift as inferred from the bTFR. The two relations are consistent at the $\approx 2\sigma$ level, with the RAR favouring no evolution and the bTFR a moderate one.

Had we instead fitted in the forward direction we would have recovered $a_1 = (-11.4 \pm 1.3) \times 10^{-10}~\mathrm{m\,s^{-2}}$ -- an $8.7\sigma$
evolution in $3.7\sigma$ tension with the RAR. That the preferred evolution
weakens substantially when the fit is conditioned on $M_{\rm bar}$ instead of $V_{\rm out}$ is a first
signature of the H{\sc i} selection we examine in
Section~\ref{hi_selection}.

Adding SPARC, the fiducial bTFR fit returns $s' = 0.276 \pm 0.006$,
$\log_{10}(V_0/\mathrm{km\,s^{-1}}) = 2.12 \pm 0.01$, and
$a_1 = (-6.3 \pm 0.9)\times 10^{-10}~\mathrm{m\,s^{-2}}$, a $6.7\sigma$ preference for evolution, opposite in sign to the corresponding RAR result
($a_1 = (+5.2 \pm 1.1)\times 10^{-10}~\mathrm{m\,s^{-2}}$), confirming a genuine tension between the two relations at the $8.2\sigma$
level. 
This disagreement is also insensitive to the fitting direction (the forward fit gives $a_1 - -7.1 \pm 0.8$, a $9.2\sigma$ tension), suggesting systematics in the SPARC data. 

The posterior distributions of the parameters from both RAR and bTFR fits for the MIGHTEE+LADUMA sample are shown in Figure~\ref{fig:rar_btfr_z_evolution_corners_mightee}; those for the combined sample with SPARC are shown in Appendix Figure~\ref{fig:rar_btfr_z_evolution_corners_combined_sample}.

\begin{figure*}
    \centering
    \includegraphics[height=6.5cm]{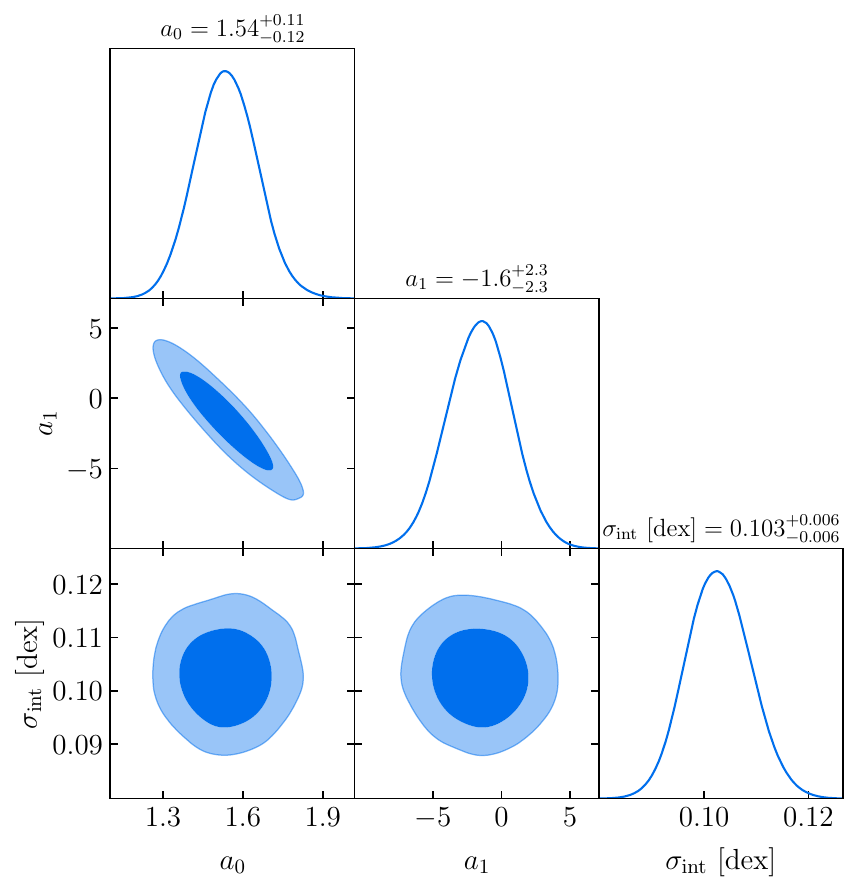}
    \hfill
    \includegraphics[height=6.5cm]{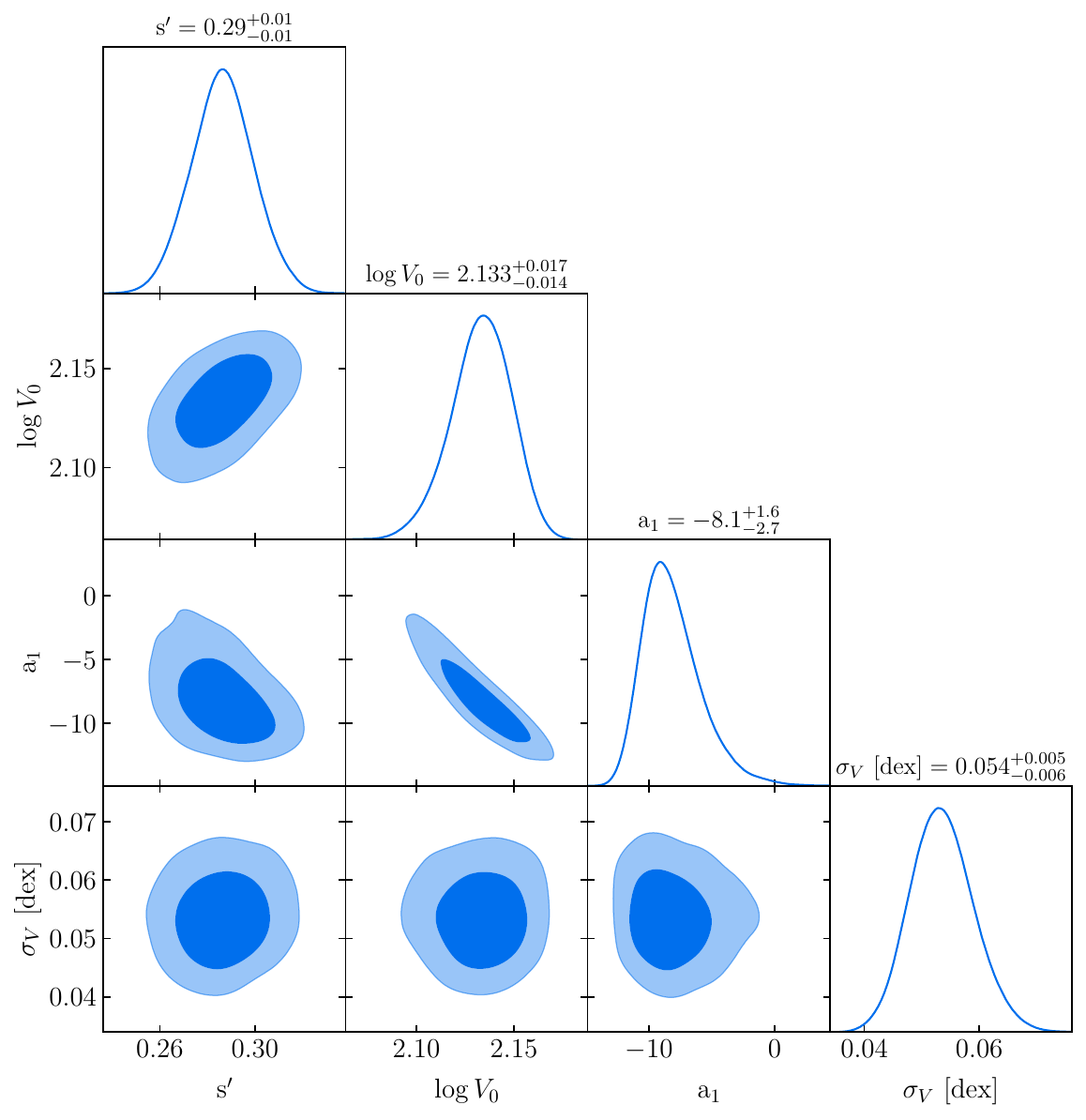}
    \caption{{\it Left}: Posterior distributions of the parameters from the
    redshift-dependent fit to the RAR for the MIGHTEE+LADUMA sample ($a_0$, $a_1$, $\sigma_{\rm int}$).
    {\it Right}: Posterior distributions of the parameters from the bTFR redshift-dependent fit (conditioned on $M_{\rm bar}$; $V_{\rm out}\,|\,M_{\rm bar}$) for MIGHTEE+LADUMA (slope, ${\rm zp}_0$, ${\rm a}_1$, $\sigma_{\rm int}$), with $a_0$ fixed to the corresponding value from the RAR fit on the left.
    Contours correspond to 68 and 95 per cent confidence intervals.}
    \label{fig:rar_btfr_z_evolution_corners_mightee}
\end{figure*}

The above parametrisation assumes the MOND relation between $a_0$ and the bTFR normalisation.
To test for evolution without this assumption, we also fit a RAR/MOND-independent form in which the zero-point evolves as a power law in $(1+z)$. Following our fiducial convention, we fit this in the inverse direction, conditioning on $M_{\rm bar}$:
\begin{equation}
    \log_{10}\!\left(\frac{V_{\rm out}}{V_0}\right) =
    s'\left[\log_{10}\!\left(\frac{M_{\rm bar}}{M_{\rm med}}\right)
    - \gamma\,\log_{10}(1+z)\right],
    \label{eq:btfr_gamma}
\end{equation}
where $\gamma > 0$ ($\gamma < 0$) indicates an increasing (decreasing) zero-point with redshift. The two parametrisations describe the same zero-point, so equating the evolution terms of Equations~(\ref{eq:zp_shift}) and (\ref{eq:btfr_gamma}) gives:
\begin{equation}
    \gamma\,\log_{10}(1+z) = -[\log_{10} a(z) - \log_{10} a_0].
\end{equation} 
Expanding to first order in z yields 
\begin{equation}
    \gamma_{\rm MOND} = -a_1/a_0
    \label{eq:btfr_gamma_a1}
\end{equation} in the small-$z \approx0 $ limit, allowing a direct comparison between the bTFR inferred evolution and RAR prediction. We again fit this model to the MIGHTEE+LADUMA sample.
\begin{figure*}
    \centering
    \includegraphics[height=6.5cm]{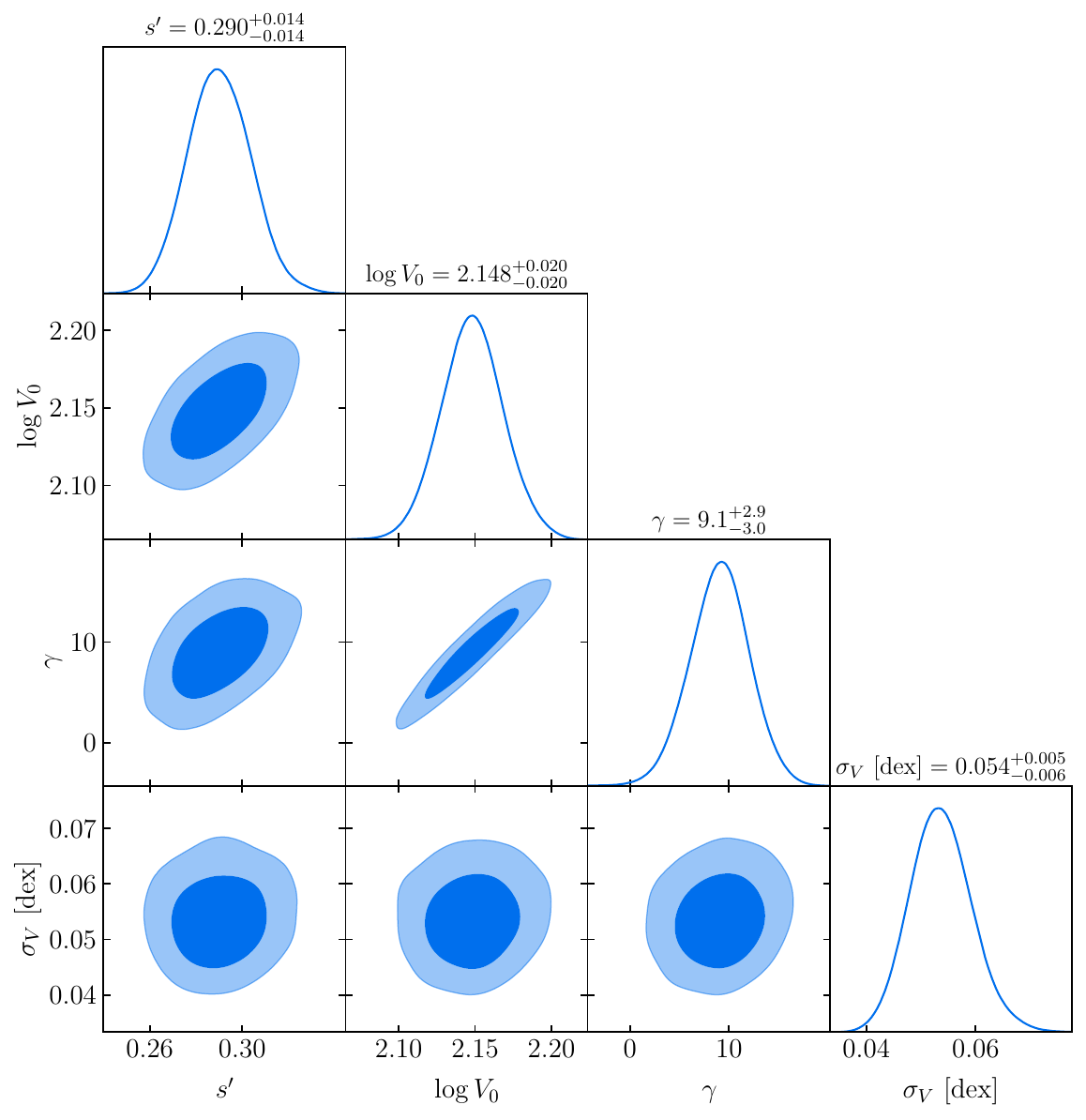}
    \hfill
    \includegraphics[height=6.5cm]{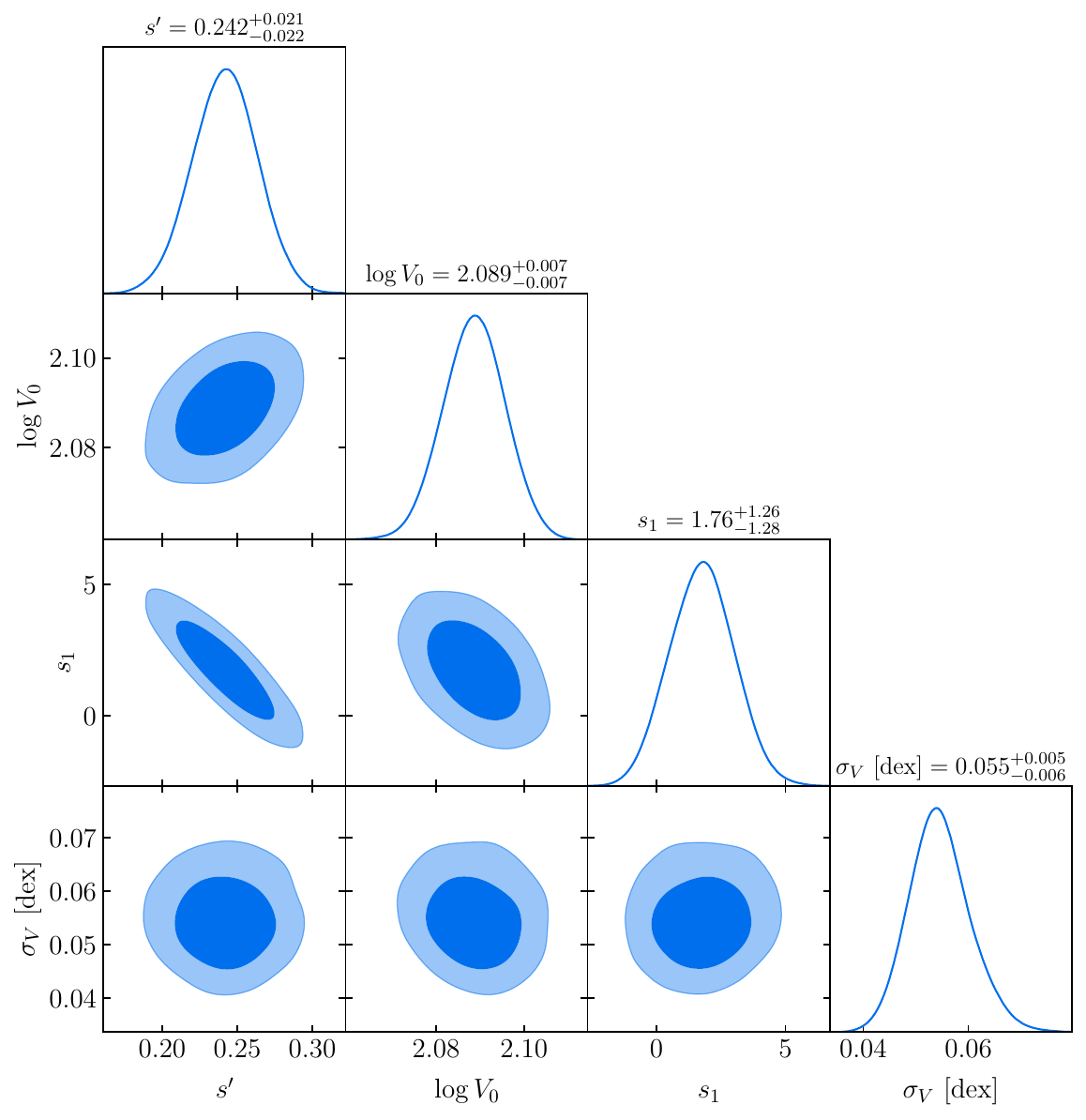}
    \caption{Posterior distributions of the parameters from the
    % RAR independent
    bTFR redshift-evolution fits (conditioned on $M_{\rm bar}$; $V_{\rm out}\,|\,M_{\rm bar}$) for the MIGHTEE+LADUMA sample.
    {\it Left}: zero-point evolution parametrised by $\gamma$ 
    (Equation~\ref{eq:btfr_gamma}).
    {\it Right}: slope evolution parametrised by $s_1$. Contours correspond to 68 and 95 per 
    cent confidence intervals.}
    \label{fig:btfr_mightee_gamma_corner}
\end{figure*}

From the MIGHTEE+LADUMA RAR redshift evolution fit ($a_0 = 1.54$, $a_1 =-1.60$, both in 
units of $10^{-10}$\,m\,s$^{-2}$), the RAR-derived redshift evolution predicts $\gamma \approx +1.0$ -- consistent with zero, since the RAR $a_1$ is itself consistent
with no evolution. From the bTFR data we instead recover 
$\gamma = 9.1 \pm 3.0$ ($3.0 \, \sigma$ from zero). As with the MOND inspired parametrisation, the bTFR prefers a moderate increase of the zero-point with redshift where the RAR predicts essentially none, now expressed through $\gamma$. A direct fit recovers a considerably stronger trend with $\gamma = 15.9 \pm 2.7$.
For completeness, the combined MIGHTEE+SPARC fit gives $\gamma = 7.5 \pm 1.4$ ($5.3\sigma$), with the corresponding MOND prediction $\gamma_{\rm MOND} \approx -4.5$; both sample combinations results are summarised in Table~\ref{tab:btfr_slope_intercept_params}. 
The posterior 
distribution of the parameters from these fits is shown in the left panel of Figure~\ref{fig:btfr_mightee_gamma_corner} for our sample (and in the left panel of Figure~\ref{fig:btfr_slope_gamma_corners_combined_sample} for the combined MIGHTEE+LADUMA+SPARC).
We additionally test for evolution in the bTFR \emph{slope} rather than its zero-point, replacing $s' \rightarrow s' + s_1\log_{10}(1+z)$ with the zero-point held fixed. In the fiducial inverse direction we find no significant slope evolution: $s_1 = 1.8 \pm 1.2$ ($1.4\sigma$) for MIGHTEE+LADUMA alone and 
$s_1 = 0.1 \pm 0.7$ for the combined sample with SPARC (Table~\ref{tab:btfr_slope_intercept_params}); a forward fit instead prefers strong slope evolution ($s_1 = -52 \pm 15$), again demonstrating the sensitivity of the evolution signal to the fitting direction, and pointing to selection effects. 

Over our limited redshift baseline a slope and zero-point evolution are in any case degenerate.
For our MIGHTEE+LADUMA sample alone the RAR shows
no significant evolution ($a_1$ consistent with zero), while the bTFR prefers a moderate one ($3.4\sigma$ from zero); the two agree in sign and remain consistent within $\approx 2\sigma$.
We caution, however, that the bTFR provides weaker redshift evolution leverage than the RAR. The RAR for our sample is fit to 476 individual data points, whereas the bTFR uses the single, outermost point per galaxy ($\sim$130 points total, Figure~\ref{fig:rar_with_outermost_point}), so the bTFR evolution parameter is less well constrained and more sensitive to the narrow redshift baseline.  

Once SPARC is added as a $z\approx0$ anchor to the RAR the redshift evolution described by $a_1$ turns significantly positive and the disagreement becomes one of opposite sign.
This inconsistency in the two relations when we add SPARC could be due to the constant mass-to-light ratio assumption in SPARC, and suggests systematics in combining samples with different selection functions and modelling choices, which can manifest as a spurious redshift trend. We investigate this in the next section.

\begin{table*}
  \centering
  \setlength{\tabcolsep}{6pt}
  \caption{% 
    Summary of redshift-evolution fits for the acceleration scale,
    parameterised as $a(z) = a_0 + a_1 z$ for two different $\Upsilon_\star$ configurations, varying and constant $\Upsilon_\star$ and both bTFR fitting directions. The inverse
    fit ($V_{\rm out}\,|\,M_{\rm bar}$) is the fiducial choice.
    $a_0$ in each bTFR fit is fixed to the $z = 0$ value from
    the corresponding RAR z-evolution fit.}
  \begin{tabular}{llccc}
    \toprule
    Sample & $\Upsilon_{\star,K}$ & Relation
           & $a_1$ [$10^{-10}$\,m\,s$^{-2}$]
           & RAR--bTFR$_{\rm inv}$ tension \\
    \midrule
    \multirow{3}{*}{MIGHTEE+LADUMA only}
      & \multirow{3}{*}{Varying}
        & RAR  & $ -1.60\pm 2.33 $ & \multirow{2}{*}{$2.0\,\sigma$} \\
      & & bTFR inverse (fiducial) & $ -8.10 \pm 2.41 $ & \\
      & & bTFR direct & $ -11.42 \pm 1.31 $ & \\
    \addlinespace[3pt]
    \multirow{3}{*}{MIGHTEE+LADUMA only}
      & \multirow{3}{*}{Constant ($\Upsilon_{\star,K} = 0.6$)}
        & RAR  & $-4.64 \pm 1.94$ & \multirow{2}{*}{${0.3\,\sigma}$} \\
       & & bTFR inverse & $-5.48\pm 2.06$ & \\
      & & bTFR direct & $-8.39 \pm 1.03$ & \\ 
    \midrule
    \multirow{3}{*}{MIGHTEE+LADUMA+SPARC}
      & \multirow{3}{*}{Varying}
        & RAR  &  $5.23 \pm 1.05 $       & \multirow{2}{*}{$8.2\,\sigma$} \\
     & & bTFR inverse & $-6.30 \pm 0.94$ & \\
      & & bTFR direct & $-7.09 \pm 0.83$ & \\
    \addlinespace[4pt]
    \multirow{3}{*}{MIGHTEE+LADUMA+SPARC}
      & \multirow{3}{*}{Constant ($\Upsilon_{\star,K} = 0.6$)}
        & RAR  & $-4.80 \pm 0.76$  & \multirow{2}{*}{$3.9\,\sigma$} \\
    & & bTFR inverse & $-8.84 \pm 0.71$ & \\
    & & bTFR direct & $-9.41 \pm 0.63$ & \\
    \bottomrule
  \end{tabular}
\label{tab:z_evolution}
\end{table*}

\subsection{Explaining the redshift trends}
\label{explaining_z_trends}
Having quantified the redshift evolution signal in Section~\ref{z_evolution}, we now examine the most likely origin of the tension between the RAR and bTFR redshift-dependent fits -- the $\Upsilon_\star$ prescription, the interpolating function (IF) shape, the velocity definition, and the H{\sc i} selection of our sample.

\subsubsection{Mass-to-light ratio variation}
First, we investigate whether the discrepancy between the RAR and bTFR results is driven by the $\Upsilon_{\star, Ks}$ prescription rather than genuine redshift cosmological evolution. We repeat both $z$ fits adopting a constant $\Upsilon_{\star, Ks} = 0.6$, matching the convention used for SPARC \citep{Lelli_2016}.
Table~\ref{tab:z_evolution} summarises the recovered redshift dependent term $a_1$ across all configurations. 
Within our MIGHTEE and LADUMA sample alone, without the SPARC anchor at $z\approx0$, the fiducial varying $\Upsilon_{\star, K}$ yields an RAR with $a_1 = -1.60\pm 2.33 \times 10^{-10}~\mathrm{m\,s^{-2}}$, but a bTFR evolution with $a_1$ = $-8.10 \pm 2.41 \times 10^{-10}~\mathrm{m\,s^{-2}}$): the two relations agree in sign, but with the RAR we find no evidence for evolution whereas the bTFR suggests a moderate one ($2.0 \sigma$ apart).
Under the constant $\Upsilon_{\star, Ks}$ assumption, the RAR-derived $a_1$ moves from consistent with zero to more negative ($a_1 = -4.64 \pm 1.94 \times 10^{-10}~\mathrm{m\,s^{-2}}$), with a tentative $2.4\sigma$ evidence for evolution, and the bTFR-derived $a_1$ becomes less negative with $a_1 = -5.48 \pm 2.06 \times 10^{-10}~\mathrm{m\,s^{-2}}$, bringing the two in agreement at the $0.3\sigma$ level, closer than the $2.0\sigma$ difference under the fiducial varying $\Upsilon_{\star, Ks}$. 

This improved agreement suggests that the larger discrepancy in the fiducial analysis arises from how the varying  $\Upsilon_{\star, K}$ prescription propagates differently through the two relations: the RAR is sensitive to the shape of the resolved stellar mass profile, while the bTFR to the total integrated mass. Under constant $\Upsilon_{\star}$, both the resolved and total mass profile use the same scaling of the $K_s$ band light, which affects both relations identically, so if there is any real trend it will be obvious in both. 
On the other hand, under the SED fitting of {\sc Bagpipes} with varying  $\Upsilon_{\star, Ks}$, the $\Upsilon_{\star}$ in each annulus is the output of a full SED fit, shaped by the age and metallicity gradients within each galaxy. More importantly, our gas rich sample prefers $\Upsilon_{\star, Ks} < 0.6$  as shown in Figure~\ref{fig:total_ml_kde}---with a median of 0.45 in $K_s$ band. This value relative to constant $0.6$ is not uniform, it varies by galaxy (see Figure~\ref{fig:total_ml_kde}) and radius (Figure~\ref{fig:ml-ratio-variations}). 

To assess whether this reflects a \emph{redshift dependence} in the mass-to-light ratio, we show the total $\Upsilon_{\star,K_s}$ as a function of redshift in the top left panel of Figure~\ref{fig:ml_gas_vs_z}. We find no significant trend (Spearman $\rho = -0.14$, $p = 0.099$): the integrated $\Upsilon_{\star,K_s}$ remains flat across our redshift range.

We further verified that the tension is not an artefact of other modelling choices. Allowing the interpolating-function (IF) shape parameter $\delta$ to vary in the RAR redshift-evolution fit 
(rather than fixing $\delta = 1$) leaves the MIGHTEE+LADUMA $a_1$ consistent with 0 ($a_1 = -0.4 \pm 2.6 \times 10^{-10}~\mathrm{m\,s^{-2}}$), suggesting that the IF shape is not the origin of the RAR--bTFR discrepancy (Figure~\ref{fig:rar_mightee_delta_z_corner}); the bTFR still prefers a moderate evolution that the RAR does not, whether $\delta$ is fixed or free. 
Although for our MIGHTEE galaxies we use $V_{\rm out}$ (the outermost measured rotation velocity), rather than $V_{\rm flat}$, any offset between the two velocity measures is unlikely to be the dominant systematics. Refitting the bTFR with $V_{\rm flat}$ only for the galaxies whose rotation curves satisfy the \citet{Lelli_2016b} flatness
criterion changes the slope by only
$\Delta s \approx 0.07$ ($\lesssim 0.3\sigma$), and shifts the zero-point by $\lesssim 0.03$~dex -- well within our quoted uncertainties, and much smaller than the difference between the forward and inverse fits.
Finally, the RAR residuals at the outermost rings---which set $V_{\rm out}$ for the bTFR---are offset from those at inner rings by only $+0.022$\,dex, far smaller than the $\sim 0.3$\,dex shift implied by the bTFR evolution we recover.

The consistency under constant $\Upsilon_{\star, K}$ does not, however, constitute evidence for genuine redshift evolution. When SPARC is included under the same $\Upsilon_{\star}$ treatment, the tension returns: $a_1 = -4.80 \pm 0.76 \times  10^{-10}~\mathrm{m\,s^{-2}}$ from the RAR and $a_1 = -8.84\pm 0.71$ from the bTFR, still a significant $3.9 \sigma$ discrepancy. A genuine cosmological evolution should manifest identically in both relations regardless of which $z \approx 0$ sample is used; the fact that SPARC anchoring reintroduces tension indicates systematics between MIGHTEE/LADUMA and SPARC samples---in baryonic modelling (spatially resolved 
versus global mass-to-light ratios, molecular gas treatment), velocity measure, or sample selection, or overall sample homogeneity.

\subsubsection{H{\sc i} sample selection}
\label{hi_selection}

\begin{figure*}
    \centering
    \includegraphics[width=0.95\textwidth]{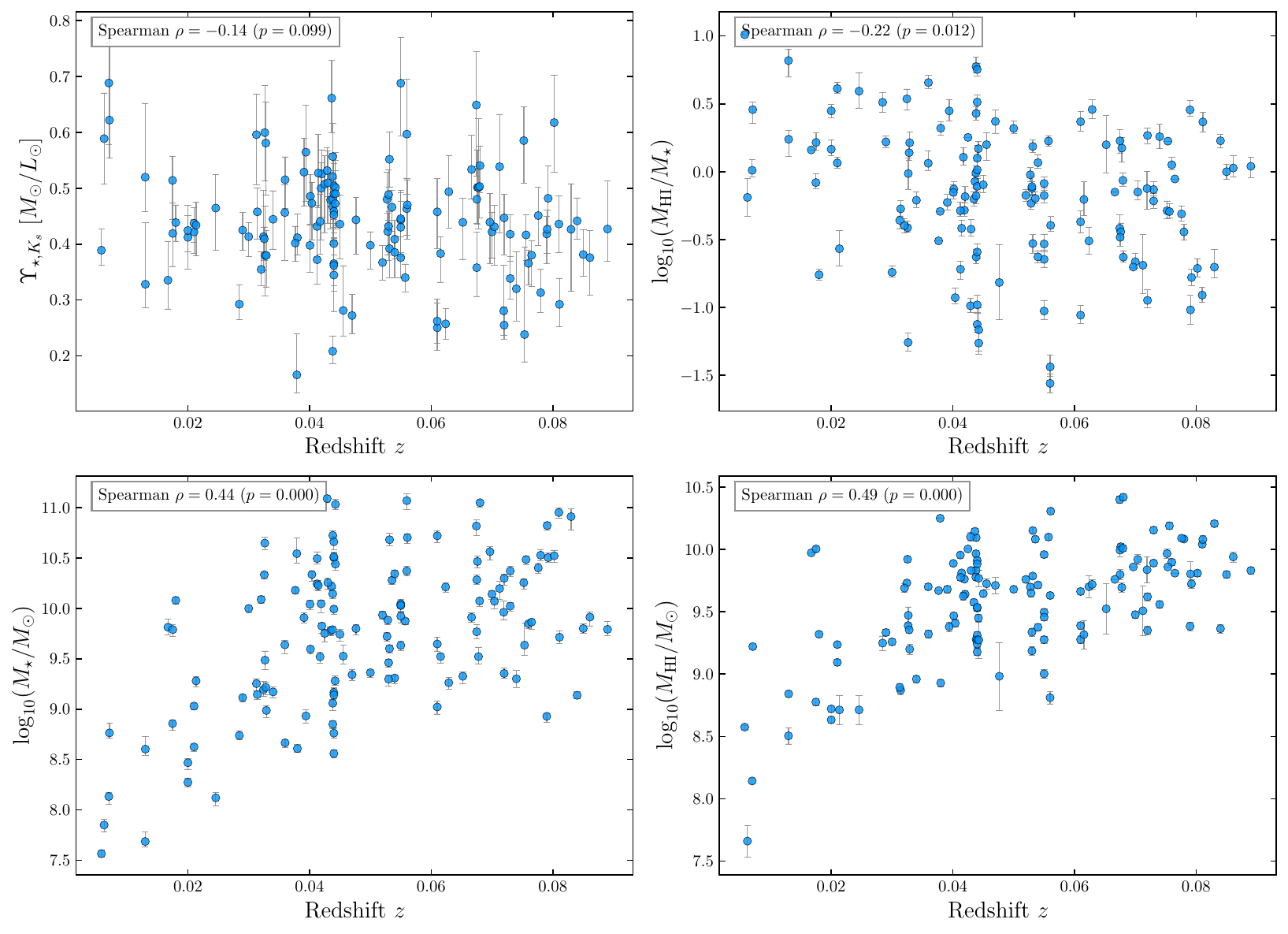}
    \caption{Total $K_s$-band stellar mass-to-light ratio $\Upsilon_{\star,K_s}$ 
(top left), atomic gas fraction $M_{\rm HI}/M_\star$ (top right), stellar mass 
(bottom left), and H{\sc i} mass (bottom right) as a function of redshift for 
the MIGHTEE+LADUMA sample. Spearman coefficients are quoted in each panel. The 
total $\Upsilon_{\star,K_s}$ shows no significant trend with redshift, while 
$M_\star$ and $M_{\rm HI}$ both rise, reflecting the selection on a flux-limited sample.}
    \label{fig:ml_gas_vs_z}
\end{figure*}

The most likely physical driver is the H\,{\sc i} selection of our sample. Both the stellar and H\,{\sc i} masses increase with redshift (Figure~\ref{fig:ml_gas_vs_z}), as expected for a flux-limited sample in which only the most massive, gas-rich systems are detected at higher redshift, coupled with the relationship between HI{\sc i} mass and stellar mass \citep[e.g.][]{Maddox2015,Sinigaglia2022,Pan2023,Pan2025}. This selection imprints more strongly on the bTFR than the RAR because of how each relation uses the baryonic mass. 

 \begin{figure*}
    \centering
    \includegraphics[width=0.95\textwidth]{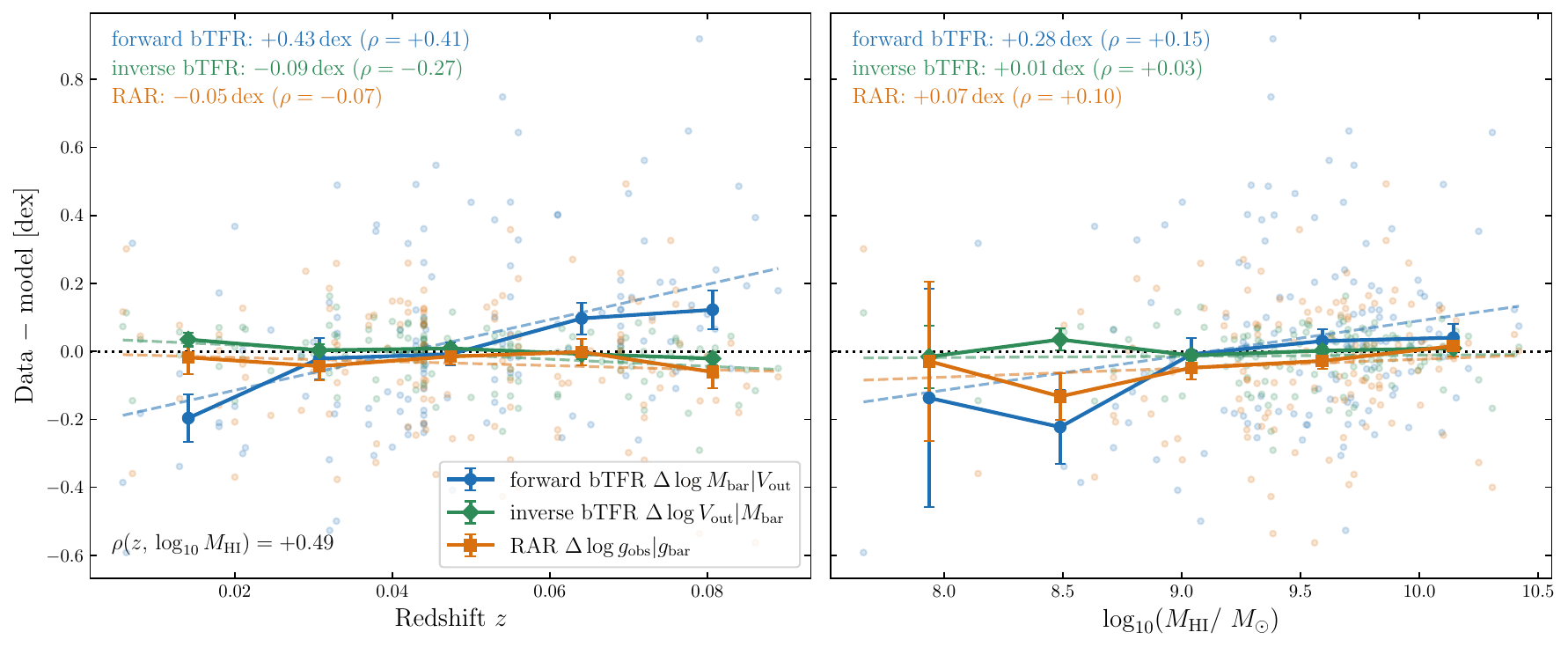}
    \caption{Vertical residuals about the \emph{non-evolving} RAR ($\Delta\log g_{\rm obs}|g_{\rm bar}$, orange) at the outermost ring, \emph{forward} bTFR ($\Delta\log M_{\rm bar}|V_{\rm out}$, blue), and the fiducial inverse bTFR ($\Delta\log V_{\rm out}|M_{\rm bar}$, green), per galaxy, against redshift (left) and the H{\sc i} mass (right). 
     The forward bTFR residual rises significantly with redshift (swing of $+0.43$\,dex, Spearman coefficient $\rho=+0.41$), whereas the RAR residual ($\Delta \log g_{\rm obs}|g_{\rm bar}$ at the outermost ring, orange) is flat ($-0.05$\,dex, $\rho=-0.07$). The selection imprint weakens significantly as $M_{\rm bar}$ moves from the ordinate (forward fit) to the abcissa (inverse fit, $\rho=-0.28$). Redshift correlates strongly with $\log M_{\rm HI}$ ($\rho=+0.49$), so that at higher $z$ only the most massive H{\sc i} systems are detected.}
    \label{fig:rar_btfr_selection_effects}
\end{figure*}

We illustrate this in Figure~\ref{fig:rar_btfr_selection_effects}, where we plot the vertical residual of each galaxy about the three \emph{non-evolving} fits -- the forward bTFR ($M_{\rm bar}\,|\,V_{\rm out}$), the fiducial inverse bTFR
($V_{\rm out}\,|\,M_{\rm bar}$), and the RAR -- against redshift and the H{\sc i}-mass.
The bTFR residual rises significantly with redshift (Spearman $\rho = +0.41$, $p<0.001$; a $+0.43$ dex change across our redshift range), whereas the inverse bTFR residual retains a
substantially weaker, but still with a significant trend (Spearman $\rho = -0.28$, $p = 0.001$); the RAR residuals on the other hand remain flat ($\rho = -0.07$, $p = 0.40$; $-0.05$ dex). 
Redshift is in turn strongly correlated with $\log M_{\rm HI}$ ($\rho = +0.49$, $p<0.001$): in our flux-limited sample we only detect the most H{\sc i}-massive systems at higher redshifts. 
Crucially, this trend appears in the relation where $M_{\rm bar}$ is the ordinate (the forward bTFR), reduced when the fit conditions on $M_{\rm bar}$ (the inverse bTFR), and absent where the baryonic mass enters only through the abscissa (the RAR). This is the signature of the H{\sc i}-mass selection rather than cosmological evolution -- a genuine evolution signal would imprint on both relations irrespective of the fitting direction. 
In the forward bTFR, $M_{\rm bar}$ is the ordinate: at higher redshift we sample only the gas rich tail -- galaxies with high  $M_{\rm bar}$ for their velocity -- so we populate a narrower slice in the bTFR, potentially mimicking the zero-point increase we recover in this case. Conditioning on $M_{\rm bar}$ instead removes some of this bias, but not all of it, so a moderate imprint survives in the inverse residuals. 
As a further quantitative check, we compare the two fitting directions for the bTFR directly. For the non-evolving relation, the forward fit with $V_{\rm out}$ as the abcissa ($M_{\rm bar}\,|\,V_{\rm out}$) returns a slope of $3.33 \pm 0.15$, whereas the fiducial inverse fit corresponds to a forward-equivalent slope of $3.73 \pm 0.15$ -- a $1.9\sigma$ separation, reflecting a non-negligible selection bias. 
The contrast is far stronger in the redshift evolution term: the forward fit ($M_{\rm bar}\,|\ V_{\rm out}$) recovers the strong zero-point evolution seen earlier ($a_1 = -11.4 \pm 1.3 \times 10^{-10}~\mathrm{m\,s^{-2}} $), whereas the inverse fit conditioned on $M_{\rm bar}$ ($V_{\rm out}\,|\,M_{\rm bar}$) substantially reduces the preferred evolution ($a_1 = -8.1 \pm 2.4 \times 10^{-10}~\mathrm{m\,s^{-2}}$), moving the bTFR $a_1$ significantly towards the RAR ($a_1 = -1.60$), with which it is consistent at $\approx 2\sigma$. 
 In the RAR, the gas instead contributes to $g_{\rm bar}$ (the abcissa), so the same bias displaces points horizontally along the relation, rather than vertically. Furthermore, the RAR $a_0$ is sensitive to the shape of the resolved stellar mass profile, so its redshift fit is tangled with the radial $\Upsilon_{\star}$ behaviour -- as evidenced by the sign flip of the RAR $a_1$ under a constant $\Upsilon_\star$, even though the integrated $\Upsilon_\star$ shows no redshift trend. 
 
 This selection bias is not confined to the high mass end. There are more low mass galaxies that are gas dominated, so at any given baryonic mass, we detect the most H{\sc i} rich galaxies due to the nature of our H{\sc i} flux limited sample. 
 The two relations are sensitive to different systematics, which is why the bTFR registers a redshift trend that the RAR does not.
 The atomic gas fraction itself shows a weak correlation with redshift (Spearman $\rho = -0.22$, $p = 0.012$, Figure~\ref{fig:ml_gas_vs_z} top right panel), consistent with competing effects of the rising gas fraction-redshift relation and the decreasing gas fraction-stellar mass relation \citep{Combes_2011, Genzel_2015, Tacconi_2018} in a flux limited sample. Robustly separating such selection effects from genuine evolution will require samples with well-characterised selection functions \citep[cf.][]{Harry_Richard_2025}.

Our findings fall within a mixed observational picture. Current bTFR measurements at $z \lesssim 0.4$ based on H{\sc i} detections show little to no evidence for evolution \citep{Ponomareva_2021, Jarvis_2025}, but are limited in sample size and large uncertainties. 
\citet{Ponomareva_2021}, using a sub-sample of our MIGHTEE galaxies at $0 < z < 0.08$ without the SPARC $z\approx 0$ anchor, found no evolution in the bTFR normalisation; their larger uncertainties are formally consistent with our results.
Our fiducial result is also in qualitative
agreement with \citet{Gogate_2023}, who find no significant evolution of the bTFR zero-point out to $z \approx 0.2$ for an H{\sc i}-selected sample with rotation velocities from carefully corrected global H{\sc i} profiles. 
At higher redshift, there are conflicting results, revealing a constant slope but evolving zero-point.
For example, \citet{Ubler_2017} find a decrease in the bTFR zero-point of $\Delta\mathrm{ZP} = -0.44$\,dex between $z \sim 0$ and 
$z \sim 0.9$, consistent with the MOND prediction derived from the RAR z evolution. 
\citet{Topal_2018} used CO and found no evidence for evolution in the slope or zero-point of the Tully-Fisher up to $z \approx 0.3$.
Cosmological hydrodynamical simulations likewise predict mild bTFR evolution: \citet{Glowacki_2021}, using the {\sc SIMBA} simulations \citep{Dave_2019}, predict a weak decrease in the intercept  between $z = 0$ and $z=1$ for most velocity measures, though notably the zero-point increases when $V_{\rm flat}$ is used specifically. 
However, the results from this work clearly demonstrate that any claims of evolution in either the RAR or the bTFR will require a careful consideration of the inherent selection biases of the samples used.

\section{Summary and conclusions} 
\label{summary}

We have presented the largest sample to date of H\,\textsc{i}-selected galaxies with both resolved H\,\textsc{i} kinematics {\em and} resolved baryonic mass profiles beyond the local Universe: 130 disc galaxies from the MIGHTEE and LADUMA surveys, reaching $z \approx 0.09$. 
For each galaxy we derived spatially resolved stellar mass surface density profiles from multi-band SED fitting allowing for a radially varying stellar mass-to-light ratio, and combined these with H\,\textsc{i} rotation curves and surface density profiles to construct both the baryonic and dynamical components of the radial acceleration relation and the baryonic Tully-Fisher Relation homogeneously. Our main conclusions are as follows

\begin{enumerate}
    \item We measure a tight RAR with an acceleration scale $a_0 = (1.50 \pm 0.05)\times 10^{-10}~\mathrm{m\,s^{-2}}$ and an intrinsic scatter of $\sigma_{\rm int} = 0.096 \pm 0.006$~dex for our fiducial direct-photometry pipeline, consistent with our previous results \citep{rar_paper_2025}, with a low intrinsic scatter supporting the interpretation of the RAR as a fundamental dynamical scaling relation.

    \item The inferred acceleration scale is robust to the choice of photometric pipeline: parametric S\'ersic fitting and non-parametric direct aperture photometry yield acceleration scales agreeing at the $0.3\sigma$ level. The intrinsic scatter is mildly sensitive to this choice, with the smoother S\'ersic profiles producing a slightly lower value by suppressing real structure that aperture photometry preserves.%2.5 sigma 

    \item Splitting the sample by photometric consistency (well-behaved galaxies for which both S\'ersic and aperture photometry agree versus less well-behaved galaxies) reveals that even when the two methods disagree, they do so in a coherent manner, shifting the inferred acceleration scale $a_0$ rather than inflating the scatter significantly. We note that galactic structure can itself affect the position along the RAR. The persistence of a tight RAR therefore reflects underlying physics, and is not an artefact of baryonic modelling. % robust to 

    \item The baryonic Tully-Fisher is recovered with a small orthogonal intrinsic scatter $\sigma_\perp \approx 0.05$~dex. We fit the relation in the inverse direction, conditioning on $M_{\rm bar}$, as the direct fit is more prone to selection bias---an important practice for all studies using flux-limited samples. The implied direct slope of the bTFR is in excellent agreement with both MIGHTEE measurements at similar redshifts \citep{Ponomareva_2021} and local SPARC-based measurements, highlighting the tight coupling between baryons and overall dynamics out to $z \approx 0.09$.

    \item Within our homogeneous MIGHTEE+LADUMA sample the RAR shows no significant evolution of the acceleration scale ($a_1 = -1.60 \pm 2.33\times10^{-10}~\mathrm{m\,s^{-2}}$). Our fiducial (inverse) bTFR fit prefers a moderate evolution ($a_1 = -8.1 \pm 2.4 \times 10^{-10}~\mathrm{m\,s^{-2}}$, $3.4\sigma$ from zero), that is nonetheless consistent with the RAR at $2\sigma$. Fitting in the forward direction instead returns $a_1 = -11.4 \pm 1.3$, a $3.7\sigma$ tension with the RAR. This dependence on fitting direction is itself a signature of selection rather than genuine evolution.

    \item The apparent evolution is not driven by a redshift trend in the integrated $\Upsilon_{\star,K_s}$ (which is flat with $z$) nor by the assumed interpolating-function shape $\delta$. Instead, our H{\sc i}-selection  preferentially detects the most gas-rich systems at higher redshift, raising $M_{\rm bar}$ at fixed $V_{\rm out}$. This selection imprints most strongly on the relation where $M_{\rm bar}$ is the ordinate (the forward bTFR), and is substantially reduced when the fit conditions on $M_{\rm bar}$ (our fiducial inverse bTFR), and is absent where the gas enters only through the abscissa $g_{\rm bar}$ (in the RAR). We confirm this directly: the forward bTFR residuals increase with redshift and with $\log M_{\rm HI}$, the inverse residuals retain only a weak trend, while the RAR residuals stays flat. Conditioning on $M_{\rm bar}$ mitigates but does not fully remove the bias, consistent with the $3.4\sigma$ preference for evolution surviving in the inverse fit. However, the strength of this signal in the inverse bTFR significantly moves towards the (lack of) evolution measured from the RAR -- both signatures of selection rather than evolution.
     Disentangling this from genuine evolution will require large samples with well-characterized selection functions.
\end{enumerate}

This work establishes a methodology for extending the RAR and bTFR beyond the local Universe using homogeneously analysed H\,\textsc{i}-selected samples with fully resolved baryonic models, and highlights the need for such samples---with robustly characterised selection functions---to disentangle systematics from genuine evolution. 
Forthcoming data from MeerKAT and ultimately the SKA will provide both the sample sizes and redshift baselines to overcome the current challenges, turning both scaling relations, the RAR and bTFR, into genuine probes to test the baryon--dynamics connection across cosmic time. We caution however that, although sample size is critical for increasing the statistical power of such measurements, results are likely to be biased unless the fundamental systematics of varying mass-to-light ratios and sample selection are robustly modelled.

\section*{Data availability}
The MIGHTEE-H{\sc i} spectral cubes are available from \href{https://archive-gw-1.kat.ac.za/public/repository/10.48479/jkc0-g916/index.html}{https://doi.org/10.48479/jkc0-g916}  \citep{Heywood_2024}. The optical and near-infrared data used to measure the stellar surface brightness properties of the galaxies are all in the public domain. Other data underlying the article are available on request to the corresponding author.

\section*{Acknowledgments}
AAV, MJJ, IH, TY, NS and RGV acknowledge the support of a UKRI Frontiers Research Grant [EP/X026639/1], which was selected by the European Research Council. MJJ, AAP and IH also acknowledge support from the STFC consolidated grants [ST/S000488/1] and [ST/W000903/1] and the Oxford Hintze Centre for Astrophysical Surveys which is funded through generous support from the Hintze Family Charitable Foundation. HD is supported by a Royal Society University Research Fellowship (grant no. 211046).
A.J.B. Acknowledges support from NSF grant AST-2308161 and from the Radcliffe Institute for Advanced Study at Harvard University. 
K.S. acknowledges support from the Natural Sciences and Engineering Research Council of Canada (NSERC). 
M.G. is supported through UK STFC Grant ST/Y001117/1. M.G. acknowledges support from the Inter-University Institute for Data Intensive Astronomy (IDIA). IDIA is a partnership of the University of Cape Town, the University of Pretoria and the University of the Western Cape. For the purpose of open access, the author has applied a Creative Commons Attribution (CC BY) licence to any Author Accepted Manuscript version arising from this submission.
The MeerKAT telescope is operated by the South African Radio Astronomy Observatory (SARAO; \url{www.ska.ac.za}), which is a facility of the National Research Foundation (NRF), an agency of the Department of Science, Technology and Innovation.
We acknowledge the use of the ilifu cloud computing facility --
\url{https://www.ilifu.ac.za}, a partnership between the University of Cape Town, the University of the Western Cape, Stellenbosch University, Sol Plaatje University and the Cape Peninsula University of Technology. The ilifu facility is supported by contributions from the Inter-University Institute for Data Intensive Astronomy (IDIA -- a partnership between the University of Cape Town, the University of Pretoria and the University of the Western Cape), the Computational Biology division at UCT and the Data Intensive Research Initiative of South Africa (DIRISA). This work made use of the IDIA processMeerKAT pipeline, developed at the Inter-University Institute for Data Intensive Astronomy (IDIA), available at
\href{https://idia-pipelines.github.io}{idia-pipelines.github.io}
(DOI: \href{https://doi.org/10.23919/URSIGASS51995.2021.9560276}{10.23919/URSIGASS51995.2021.9560276}).

% For the purpose of open access, the author has applied a Creative Commons Attribution (CC BY) licence to any Author Accepted Manuscript version arising from this submission. HD: Not needed now that MNRAS is fully open access.
We thank Kyle Oman, Pavel E. Mancera Pi$\mathrm{\tilde{n}}$a, Sambatriniaina Rajohnson, Simon B. De Daniloff for comments on the manuscript.
This work is based on observations made with the MeerKAT telescope.
This research made use of {\sc Photutils}, an Astropy package for
detection and photometry of astronomical sources \citep{photutils}, {\sc sep} \citep{Bertin_1996}, {\sc Roxy} \citep{roxy}, {\sc Scipy} \citep{Scipy_2020}, {\sc fgivenx} \citep{fgivenx}, {\sc numpyro} \citep{numpyro1, numpyro2}.
%%%%%%%%%%%%%%%%%%%%%%%%%%%%%%%%%%%%%%%%%%%%%%%%%%

%%%%%%%%%%%%%%%%%%%% REFERENCES %%%%%%%%%%%%%%%%%%

% The best way to enter references is to use BibTeX:

\bibliographystyle{mnras}
\bibliography{ref.bib} 
% Alternatively you could enter them by hand, like this:
% This method is tedious and prone to error if you have lots of references
%\begin{thebibliography}{99}
%\bibitem[\protect\citeauthoryear{Author}{2012}]{Author2012}
%Author A.~N., 2013, Journal of Improbable Astronomy, 1, 1
%\bibitem[\protect\citeauthoryear{Others}{2013}]{Others2013}
%Others S., 2012, Journal of Interesting Stuff, 17, 198
%\end{thebibliography}

%%%%%%%%%%%%%%%%%%%%%%%%%%%%%%%%%%%%%%%%%%%%%%%%%%

%%%%%%%%%%%%%%%%% APPENDICES %%%%%%%%%%%%%%%%%%%%%

\appendix
%\section{Sample Data Table}
%\input{gbar_gobs_table}
\label{appendix}
% table with sample properties maybe and some ML variations 
\section{RAR and bTFR evolution}
\label{btfr_evolution}
\begin{figure*}
    \centering
    \includegraphics[height=6.5cm]{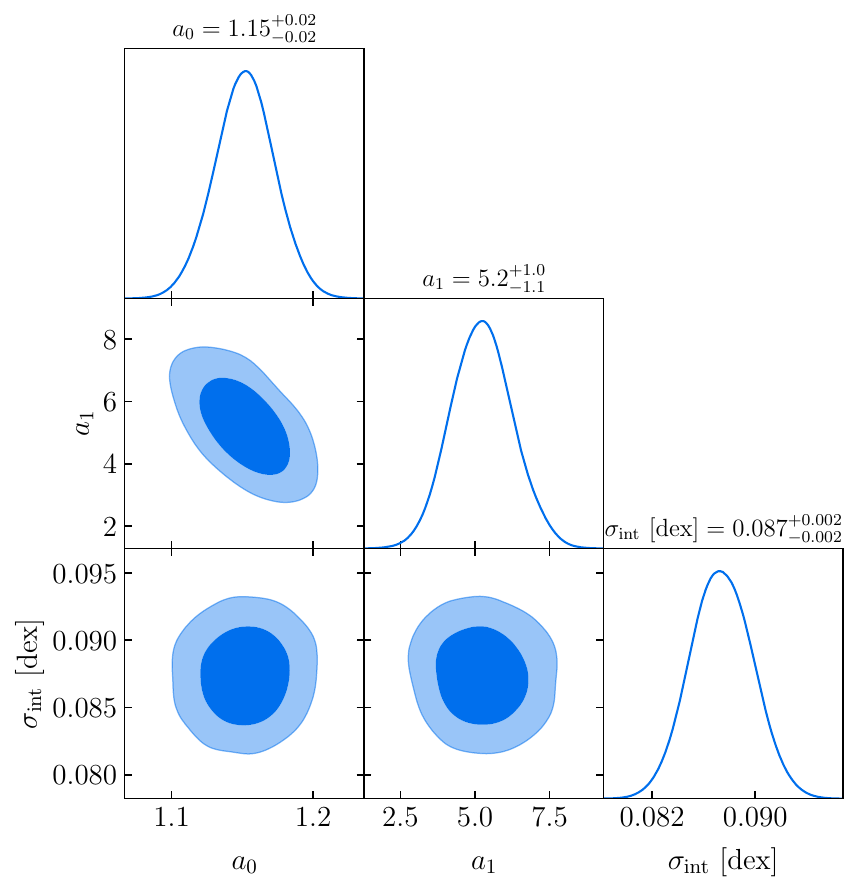}
    \hfill
    \includegraphics[height=6.5cm]{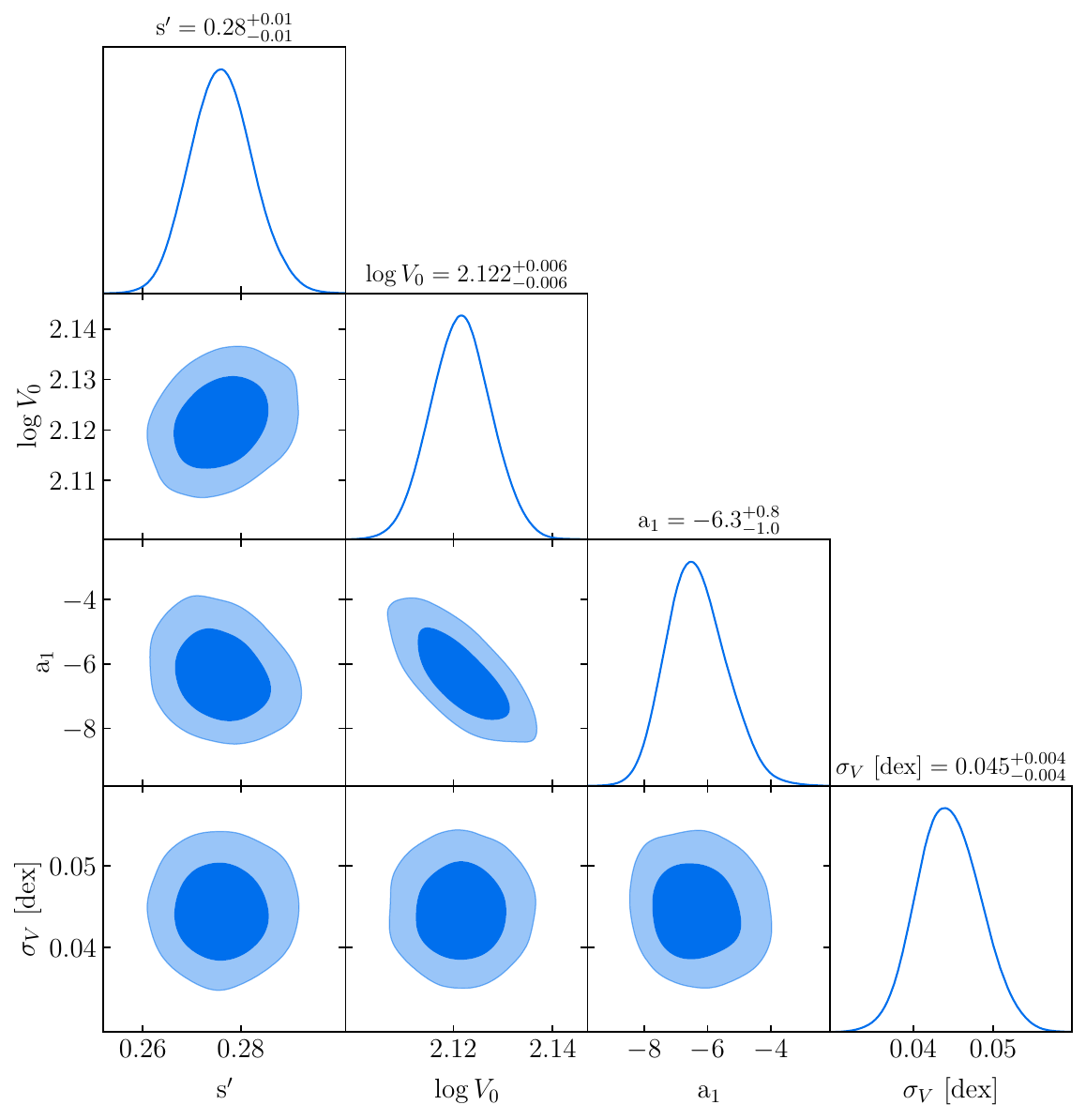}
    \caption{{\it Left}: Posterior distributions of the parameters from the
    redshift-dependent fit to the RAR ($a_0$, $a_1$, $\sigma_{\rm int}$) for the combined sample with SPARC.
    {\it Right}: Posterior distributions of the parameters from the redshift-dependent fit to
    the bTFR conditioned on $M_{\rm bar}$ ($V_{\rm out}\,|\,M_{\rm bar}$) for the combined sample with SPARC. $a_0$ is fixed to the corresponding value from the RAR fit on the left.
    Contours correspond to 68 and 95 per cent confidence intervals.}
    \label{fig:rar_btfr_z_evolution_corners_combined_sample}
\end{figure*}

% first table
\begin{table*}
  \centering
  \caption{Model-independent bTFR redshift-evolution fits using the
    parametrisation of Equation~(\ref{eq:btfr_gamma}) and
    $s \rightarrow s + s_1\,\log_{10}(1+z)$. All fits use the fiducial varying
    $\Upsilon_{\star,K}$. The MOND predictions $\gamma_{\rm MOND}=-a_1/a_0$ are
    derived from the corresponding RAR $z$-fits. Slope and zero-point evolution
    cannot be constrained simultaneously over our redshift baseline, so $\gamma$
    and $s_1$ are fit separately.}
  \begin{tabular}{llcc}
    \toprule
    Sample & Parameter & Fitted value & MOND prediction \\
    \midrule
    \multirow{2}{*}{MIGHTEE+LADUMA}
      & $\gamma$ & $9.06 \pm 3.02\ (3.0\sigma)$ & $+1.05$ \\
      & $s_1$    & $1.76 \pm 1.24\ (1.4\sigma)$ & --- \\
    \addlinespace[4pt]
    \multirow{2}{*}{MIGHTEE+LADUMA+SPARC}
      & $\gamma$ & $7.54 \pm 1.41\ (5.3\sigma)$ & $-4.55$ \\
      & $s_1$    & $0.10 \pm 0.73\ (0.1\sigma)$ & --- \\
    \bottomrule
  \end{tabular}
  \label{tab:btfr_slope_intercept_params}
\end{table*}

% then plots 
\begin{figure*}
    \centering
    \includegraphics[width=0.45\textwidth]{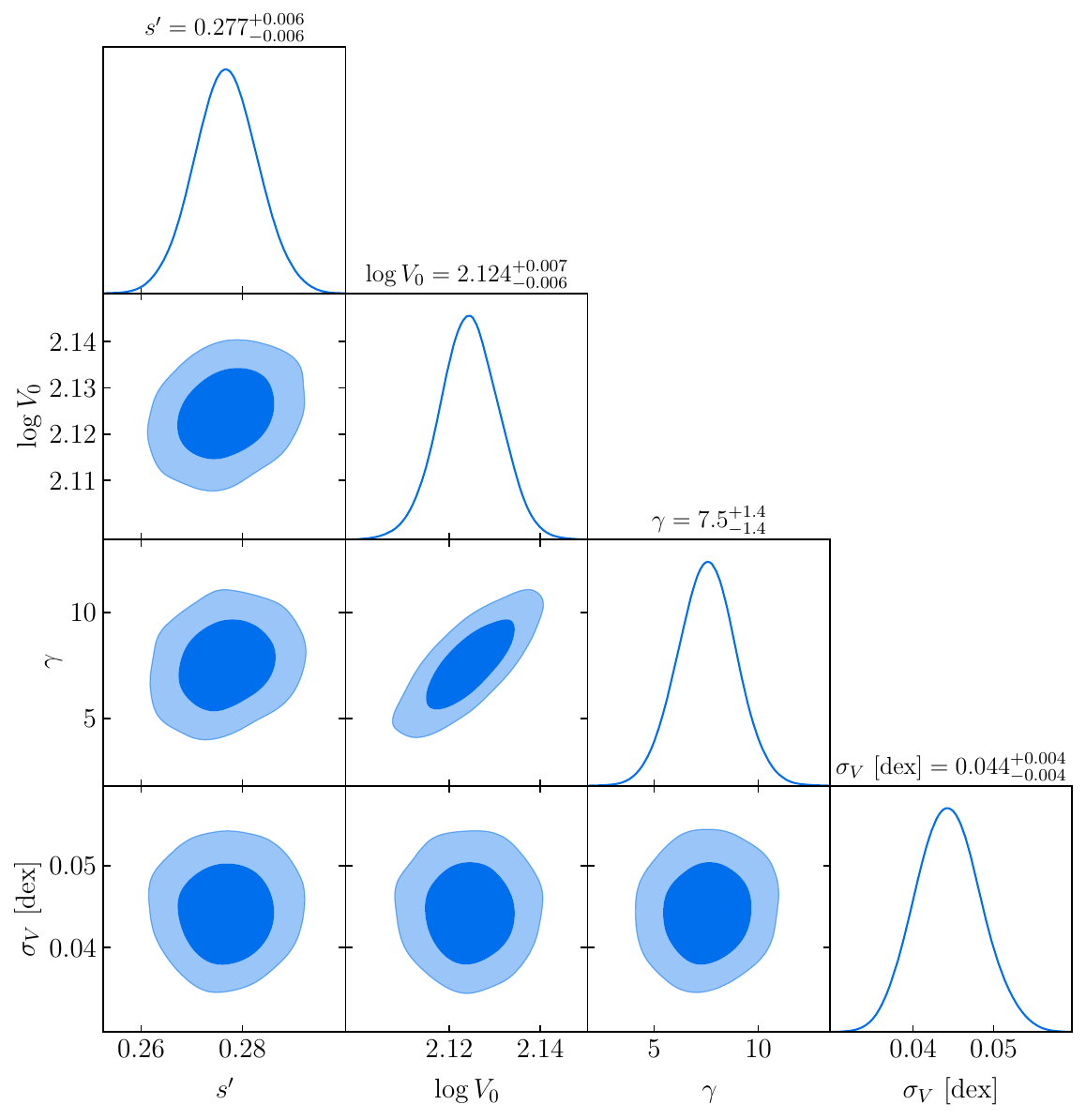}
    \hfill
    \includegraphics[width=0.45\textwidth]{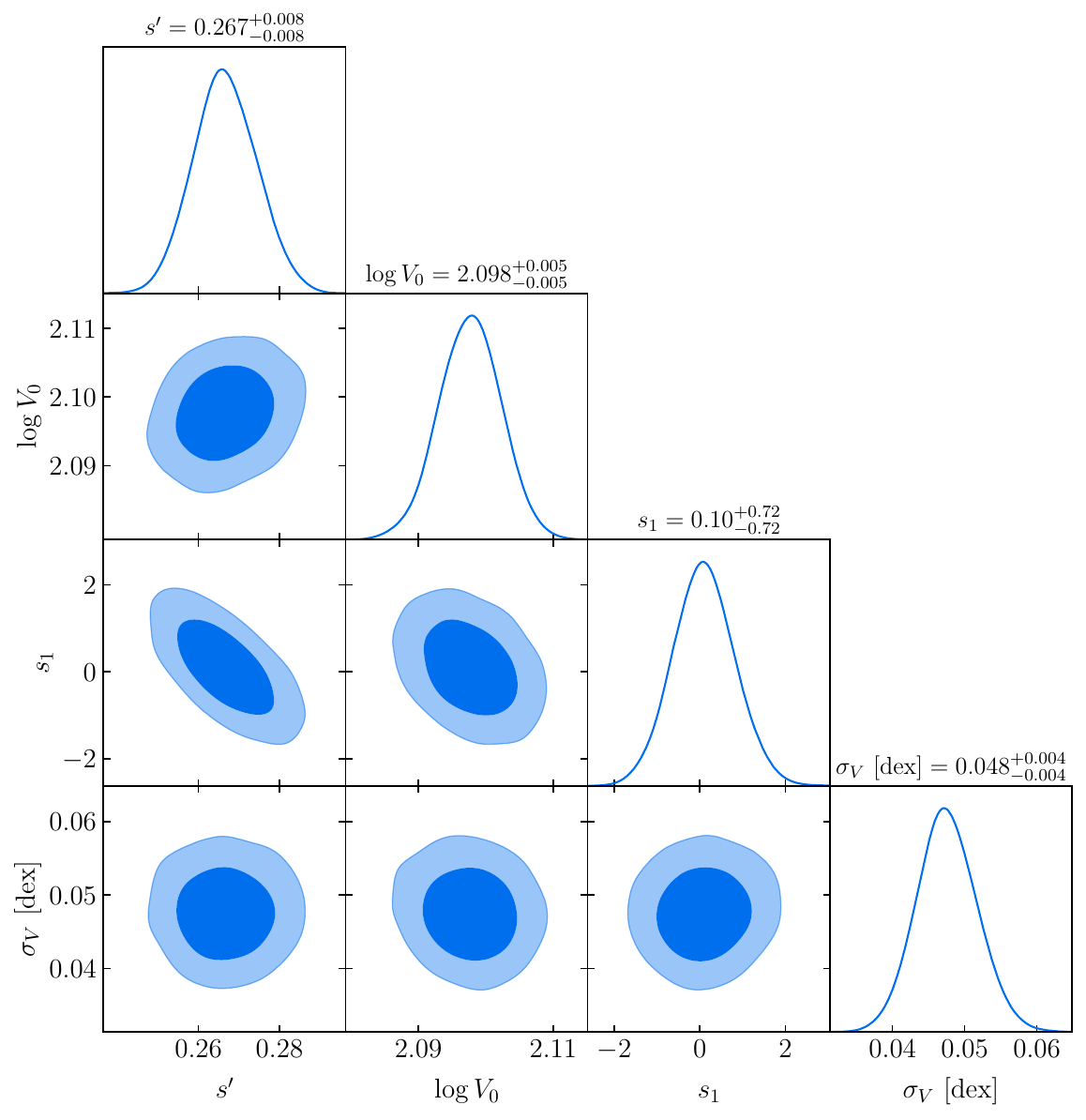}
    \caption{Posterior distributions of the parameters from the RAR-independent bTFR redshift-evolution fits for the MIGHTEE+LADUMA+SPARC sample.
    {\it Left}: Zero-point evolution parametrised by $\gamma$ 
    (Equation~\ref{eq:btfr_gamma}).
    {\it Right}: Slope evolution parametrised by $s_1$. Contours correspond to 68 and 95 per 
    cent confidence intervals.}
    \label{fig:btfr_slope_gamma_corners_combined_sample}
\end{figure*}

\begin{figure}
    \centering
    \includegraphics[width=0.5\textwidth]{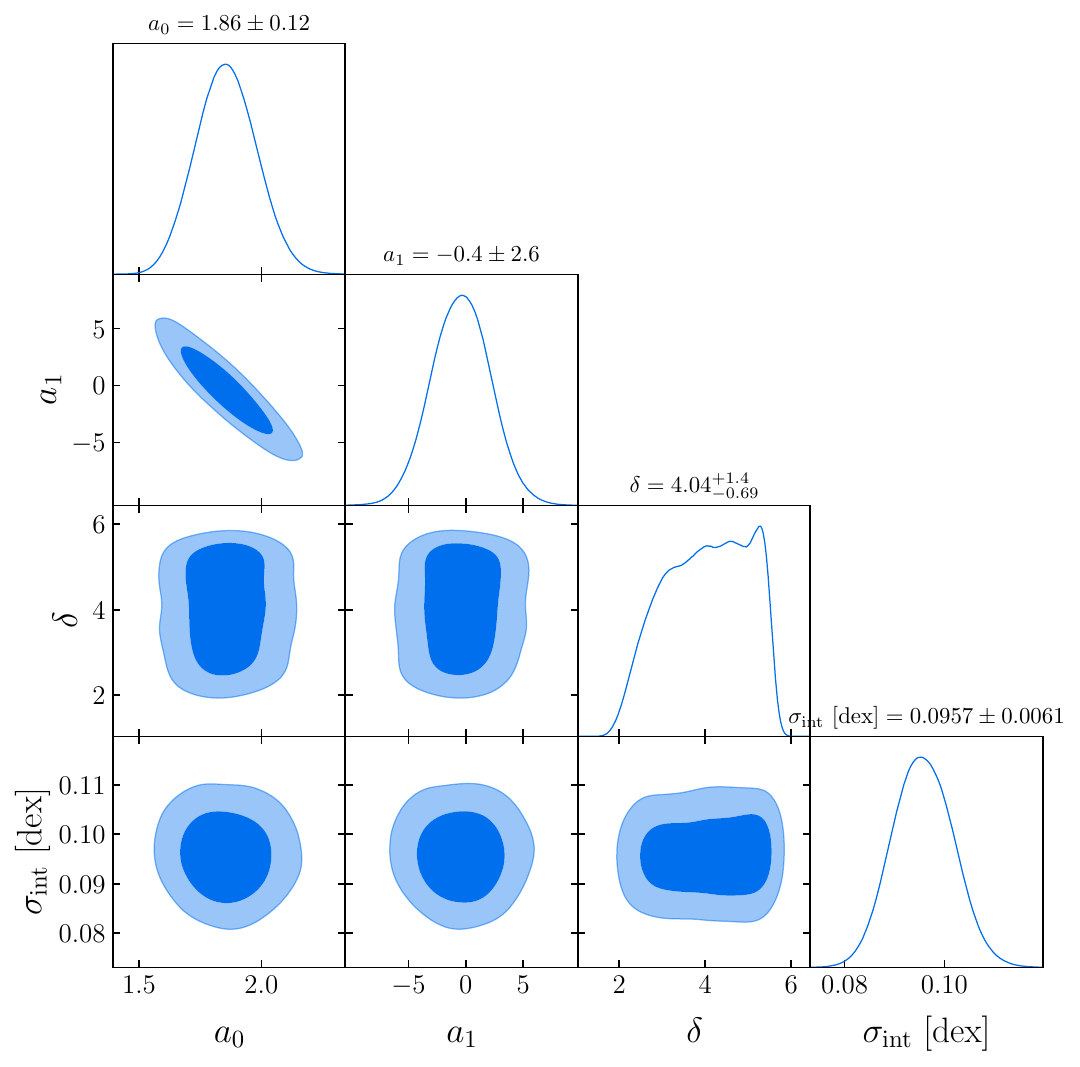}
    \caption{Posterior distribution of parameters from the RAR z dependent fit with the shape parameter $\delta$ (Equation~\ref{eq:delta_family} with the z dependence in the acceleration scale) free.}
    \label{fig:rar_mightee_delta_z_corner}
\end{figure}

%\section
%%%%%%%%%%%%%%%%%%%%%%%%%%%%%%%%%%%%%%%%%%%%%%%%%%
% Don't change these lines
\bsp	% typesetting comment
\label{lastpage}
\end{document}